\documentclass[aps,notitlepage,showpacs,superscriptaddress,reprint, longbibliography]{revtex4-1}

\usepackage{hyperref}
\usepackage{placeins}
\usepackage{graphicx}
\usepackage{mhchem}   %Names of chemical elements (typesetting)
\usepackage{mathrsfs}		%A nice typesetting for units
\usepackage[per-mode = symbol]{siunitx}			
\usepackage{gensymb}
\usepackage[T1]{fontenc}	
\usepackage[english]{babel}
\usepackage{blindtext} 
\usepackage{xcolor}
\usepackage{array}
\usepackage{float}
\newcolumntype{H}{>{\setbox0=\hbox\bgroup}c<{\egroup}@{}}
\newcolumntype{Z}{>{\setbox0=\hbox\bgroup}c<{\egroup}@{\hspace*{-\tabcolsep}}}
\usepackage[margin=1.1in]{geometry}
\DeclareSIUnit\erg{erg}

\begin{document}
\title{\LARGE Light-driven ultrafast phonomagnetism}
%\title{\LARGE Resonant lattice control of competing magnetic order}

%\date{\today}
\author{D.~Afanasiev}
\email{equal contribution}
%\email{Contact e-mail: d.afanasiev@tudelft.nl}
\affiliation{Kavli Institute of Nanoscience,
	Delft University of Technology, P.O. Box 5046, 2600 GA Delft, The Netherlands}
\author{J.~R.~Hortensius}
\email{equal contribution}
\affiliation{Kavli Institute of Nanoscience,
	Delft University of Technology, P.O. Box 5046, 2600 GA Delft, The Netherlands}
\author{B.~A.~Ivanov}
\affiliation{Institute of Magnetism, National Academy of Sciences and Ministry of Education and Science, 03142 Kiev, Ukraine}
\affiliation{National University of Science and Technology \guillemotleft MISiS\guillemotright, Moscow, 119049, Russian Federation}
\author{A.~Sasani}
\affiliation{CESAM QMAT Physique Th\'eorique des Mat\'eriaux, Universit\'e de Li\`ege,  B-4000 Sart Tilman, Belgium}
\author{E.~Bousquet}
\affiliation{CESAM QMAT Physique Th\'eorique des Mat\'eriaux, Universit\'e de Li\`ege,  B-4000 Sart Tilman, Belgium}
\author{Y.~M.~Blanter}
\affiliation{Kavli Institute of Nanoscience,
	Delft University of Technology, P.O. Box 5046, 2600 GA Delft, The Netherlands}
\author{R.~V.~Mikhaylovskiy}
\affiliation{Department of Physics, Lancaster University, Bailrigg, UK}
\author{A.~V.~Kimel}
\affiliation{Radboud University Nijmegen, Institute for Molecules and
	Materials, 6525 AJ Nijmegen, The Netherlands}
\affiliation{Moscow Technological University, MIREA, Vernadsky Ave. 78, 119454, Moscow, Russian Federation}
\author{A.~D.~Caviglia}
\affiliation{Kavli Institute of Nanoscience,
	Delft University of Technology, P.O. Box 5046, 2600 GA Delft, The Netherlands}

\begin{abstract}
	Exciting atomic oscillations with light is a powerful technique to control the electronic properties of materials, leading to remarkable phenomena such as light-induced superconductivity and ultrafast insulator to metal transitions. Here we show that light-driven lattice vibrations can be utilised to encode efficiently spin information in a magnetic medium. Intense mid-infrared electric field pulses, tuned to resonance with a vibrational normal mode of antiferromagnetic DyFeO$_3$, drive the emergence of long-living weak ferromagnetic order. Light-driven phonon displacements promptly lower the energy barrier separating competing magnetic states, allowing the alignment of spins to occur within a few picoseconds, via non-equilibrium dynamics of the magnetic energy landscape.
\end{abstract}

\maketitle

The ability to encode spin information with minimal energy dissipation is of great scientific and technological interest~\cite{Wolf2001}. In magnetoelectric materials, controlling spins via strain and electric fields becomes viable, providing an energy-efficient route towards information recording~\cite{Mundy2016,Manipatruni2018,Schlom2014,Spaldin2019}. Recently, selective excitation of elementary vibrations of the crystal lattice (phonons) by ultrashort pulses of light has emerged as a novel, low-energy route to control material properties, extending strain engineering to the ultrafast timescales~\cite{Rini2007,Fausti2011,Mankowsky2014,Kubacka2014,Mitrano2016,Subedi2014,Knap2016}. Although pioneering experiments have demonstrated that driving optical phonons can affect magnetism~\cite{Nova2017,Maehrlein2018}, no switching of the spin orientation and light-induced magnetic symmetry breaking has been shown so far. Ultrafast magnetic switching is a thriving field but its main focus until now has been the excitation of spin dynamics via electronic absorption~\cite{Schlauderer2019,Stupakiewicz2017}. Here we investigate light-induced magnetism in dysprosium orthoferrite~(DyFeO$_3$), a magnetoelectric material with one of the strongest interactions between spins and the crystal lattice~\cite{Tokunaga2008,Nakajima2015}. We show that a sub-ps pulse of an intense mid-infrared electric field, tuned to resonance with a phonon mode, drives a coherent spin-reorientation within a half-cycle of the spin precession, developing long-living weak ferromagnetic order. Light-induced magnetism emerges via a non-equilibrium metastable route, inaccessible via a thermodynamic transformation. We show that phonomagnetism originates from ultrafast lattice control of the anisotropic exchange interaction.

\begin{figure*}
	\centering
	\includegraphics[width =\linewidth]{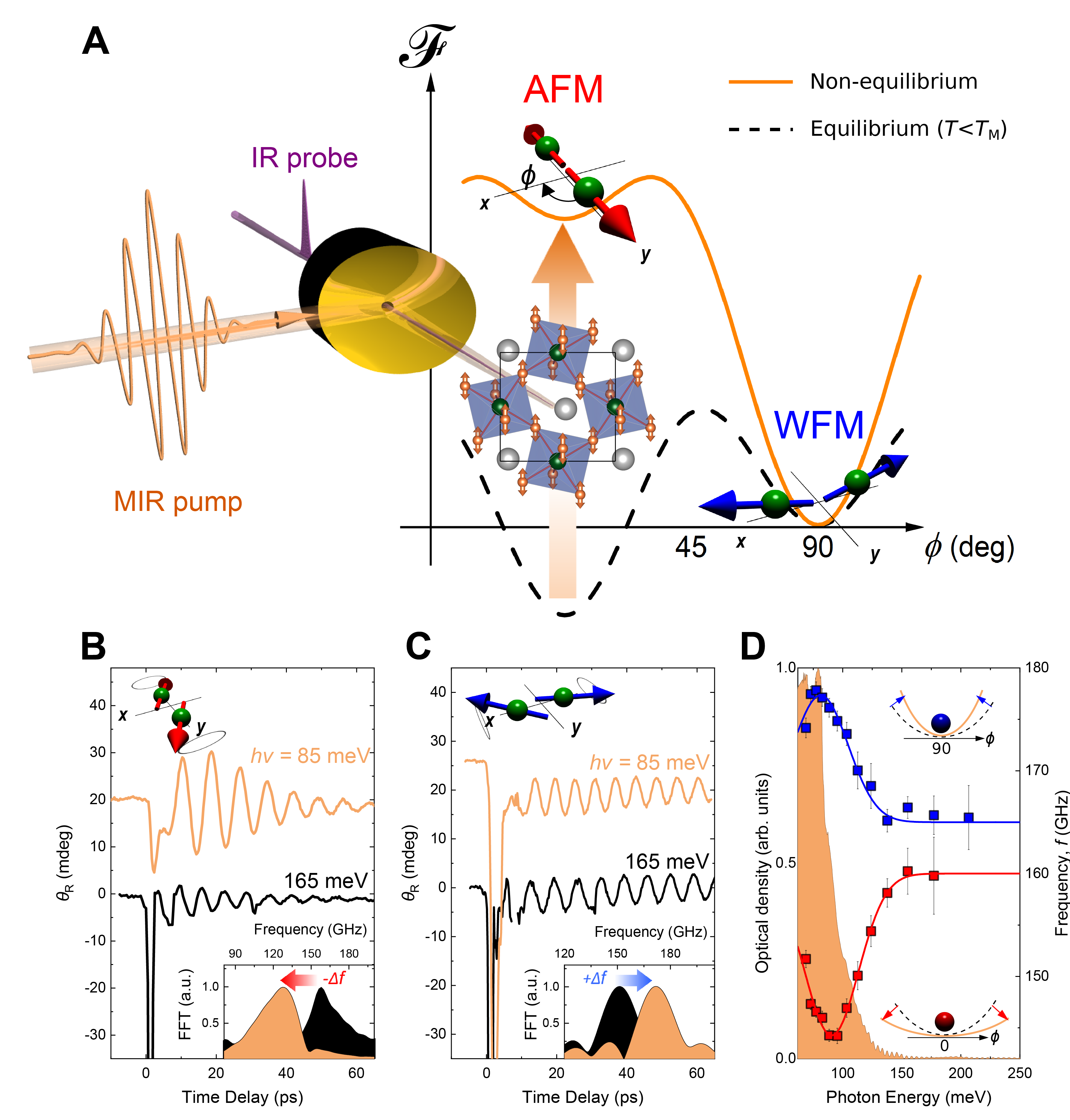}
	\caption{\textbf{Phonon-driven reconfiguration of the magnetic potential.} (\textbf{A}) % The Fe$^{3+}$ spin arrangements in the AFM phase ($T < T_\textnormal{M}$) and WFM phase ($T > T_\textnormal{M}$). %Purple/orange spheres: \ce{O^2-}/\ce{Fe^3+} ions. %Dysprosium ions are not shown, for clarity. 
		%The antiferromagnetically ordered \ce{Fe^3+} spins form two magnetic sublattices denoted \textbf{S}$_1$ and \textbf{S}$_2$ and are oriented in the $y$- and $x$-direction in the AFM and WFM phase respectively. %Above $T_\textnormal{M}$=\SI{51}{K} the \ce{Fe^3+} spins cant partially in the $z$-direction, which results in a net magnetic moment. 
		Schematics of the magnetic potential $\mathscr{F}$ as a function of the angle $\phi$ the spins form with the $y$-axis before (dashed black) and after (solid orange) phonon excitation. For simplicity the schematics is shown only for the AFM phase ($T<T_\textnormal{M}$). The magnetic configurations corresponding to the AFM and WFM phases are depicted by the red and blue arrows situated in the $xy$-plane. The orange arrow highlights the destabilization of the AFM ground state together with the reduction of the kinetic barrier separating the phases. The eigenmode of the laser-excited $B_\textnormal{1u}$ lattice vibration is shown schematically. The orange/green/grey spheres are \ce{Fe^3+}/\ce{O^2-}/\ce{Dy^3+} ions. The experimental set-up is depicted by the optical path of the linearly polarized mid-infrared pump and infrared probe pulses. The pump pulse is reflected from the parabolic mirror and steered to the sample.
		%representation of the ultrafast evolution of the equilibrium magnetic energy landscape (black dashed line) corresponding to a first order phase transition into a transient non-equilibrium landscape (yellow line) after the phonon excitation. This reconfiguration promotes a coherent transition to the other magnetic state. %The transient potential gives rise to normal mode oscillations with a different frequency from equilibrium, determined by the potential curvature in the minimum.
		(\textbf{B} and \textbf{C}) Time-resolved transient rotation of the probe polarization plane after light excitation with photon energy of \SI{85}{\milli\electronvolt} (orange traces) and \SI{165}{\milli\electronvolt} (black traces) performed in the AFM (B) and WFM (C) phases. The bottom insets show the FFT spectra of the soft mode oscillations. The top insets are schematics of the corresponding spin precessions. (\textbf{D}) Central frequency of the excited soft mode as a function of the photon energy of the pump pulse in the two magnetic phases. The solid lines serve as a guide to the eye.  %The red and blue squares correspond to the AFM ($T = $ \SI{35}{\kelvin}) and WFM ($T = $ \SI{55}{\kelvin}) phases, respectively. 
		The background, shaded, curve shows the sample's optical density measured using FTIR absorption spectroscopy.  %\textcolor{red}{The pump fluence was ..}
		%The landscape $\mathscr{F}$ is drawn as function of projection of the iron spin \textbf{S}=$\left\{S_\text{x},S_\text{y}\right\}$ on the in-plane crystallographic axes. The angle $\phi$ is the order parameter of the transition.
	}
	\label{Figure1}
\end{figure*}

\begin{figure*}
	\centering
	\includegraphics[width =\linewidth]{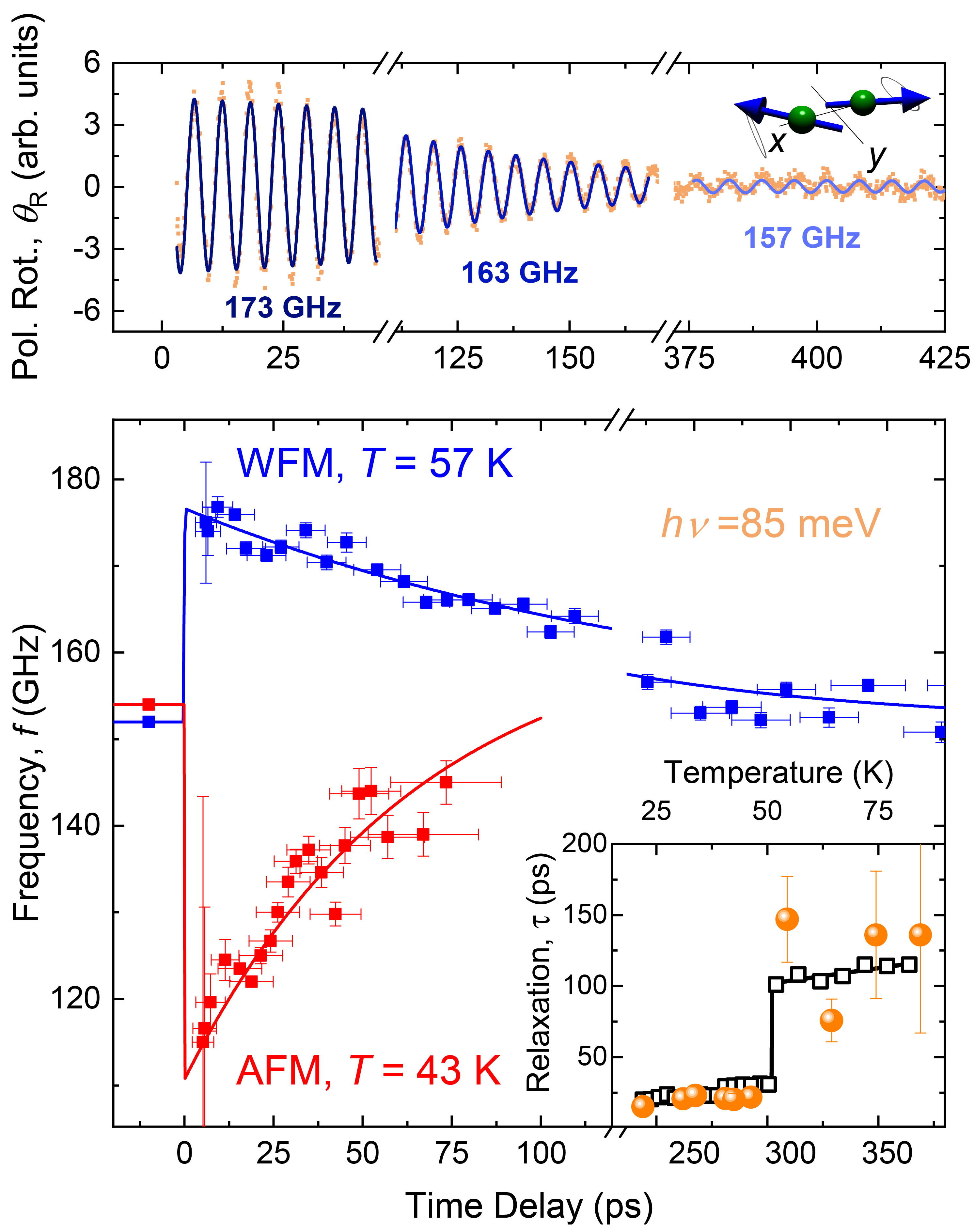}
	\caption{\textbf{Ultrafast dynamics of the soft mode frequency.} (Top panel) Trace of the spin precession dynamics excited in the WFM phase.~The solid lines are sine fits to extract a time varying frequency $f$.~(Bottom panel) The resulting dynamics of the frequency $f$ as measured in the AFM (red) and WFM (blue) phases. The solid lines are guides to the eye. Inset: Relaxation time of the frequency across $T_\textnormal{M}$ (yellow markers) imposed on the decay time of the spin precession (black markers).
		%	Time-resolved dynamics of the spin precession frequency $f$ as measured in the AFM (red) and WFM (blue) phases after the phonon-pumping. The solid lines are guides to the eye.  Inset: Relaxation time of the frequency shift across $T_\textnormal{M}$ (yellow markers) imposed on the decay time of the spin precession.%, obtained after off-resonant excitation (\SI{165}{\milli\electronvolt}). The solid line serves as guide to the eye %\textcolor{red}{The pump fluence was ..} 
	} 
	\label{Figure2}
\end{figure*}

The magnetic insulator \ce{DyFeO3} crystallizes in a perovskite orthorhombic structure and exhibits antiferromagnetic spin order set by the Fe-Fe isotropic and the Fe-Dy anisotropic exchange interactions~\cite{Yamaguchi1974}. The anisotropic exchange results in the spin reorientation transition, in which the direction of the magnetic easy-axis for the iron spins changes between $y$- and $x$-crystal axis upon crossing the Morin point ($T_\textnormal{M}$=51~K)~\cite{Afanasiev2015}. The reorientation is accompanied by a change in magnetic symmetry, from a purely antiferromagnetic state (AFM), to a weakly ferromagnetic state (WFM) characterized by the emergence of a net magnetization $M$ (supplementary text S2). The Morin transition in DyFeO$_3$ is of the first-order and is heavily influenced by the nucleation processes resulting in a very narrow temperature hysteresis. 

The spin-reorientation is described by the magnetic anisotropy energy $\mathscr{F}$ which depends on the temperature and the angle $\phi$~the spins form with the $y$-axis~\cite{Zvezdin1979}. In a broad temperature range this potential features two characteristic minima at $0^{\circ}$ and $90^{\circ}$ (see Fig.~\ref{Figure1}) signaling two ordered states~\cite{Afanasiev2016}. The height of the kinetic barrier separating the two competing states, as well as their relative energy, is controlled by the strength of the Fe-Dy exchange. This interaction is strongly sensitive to changes in the crystal environment, both via direct modulation of the Fe-Dy electronic overlap and changes in the orbital state of the rare-earth $4f$ multiplet~\cite{Khim2011,Baierl2016}. In our study, we consider light-induced oscillations of atoms driven far from their equilibrium positions to manipulate the magnetic energy landscape~\cite{Mankowsky2016,Fechner2018} in order to realize the lattice control of the spin arrangement on the ultrafast timescale.

In our experiments we investigate spin dynamics in single crystals of $z$-cut \ce{DyFeO3}. The light-excitation is provided by an intense (electric field in excess of~10~MV/cm), phase-locked mid-infrared impulsive source (250~fs,~\SI{1}{\kilo\hertz}), tunable in a broad photon energy range from \SIrange[range-units = single, range-phrase = {~to~}]{65}{250}{\milli\electronvolt}. This energy range covers the broad absorption peak of the highest frequency B$_\textnormal{1u}$ infrared-active phonon mode, associated with periodic stretching of the \ce{Fe}-\ce{O} bonds, see Fig.~\ref{Figure1}a and Fig.~S4,S5. The ensuing spin dynamics is measured, in a conventional pump-probe scheme, by tracking the polarization rotation, imprinted by the magneto-optical Faraday effect, on co-propagating near-infrared probe pulses at the photon energy of \SI{1.5}{\electronvolt}. In our experimental geometry, the Faraday rotation probes the magnetization dynamics along the normal direction. % (see supplementary information). 
Importantly, the frequency $f$ of the antiferromagnetic spin precession mode~(the soft mode of the Morin phase transition), informs us on the local curvature of the magnetic potential, according to the relation $f^2\sim\frac{\partial^2 \mathscr{F}}{\partial \phi^2}|_{\phi=\phi_0}$, both in the AFM~($\phi_0=0^{\circ}$)~and~WFM~($\phi_0=90^{\circ}$)~phases~\cite{Yusupov2010,Yamaguchi2015}.

%infrared
% centered at \SI{65}{\milli\electronvolt}

Figure~\ref{Figure1}c and \ref{Figure1}d show light-induced dynamics of the Faraday signal in the AFM and WFM phases revealing coherent oscillations, corresponding to spin precession around their respective equilibria. These dynamics are induced by pump pulses tuned in resonance with the lattice vibrational mode (85 meV). As a control experiment, we excite magnon oscillations via impulsive stimulated Raman scattering (ISRS)~\cite{Kalashnikova2007,Afanasiev2016} using pulses (165 meV) tuned away from lattice or electronic resonances. These conventional time-domain spectroscopy measurements serve as a probe of the curvature of the equilibrium potential.%and are consistent with previous reports~.%, in the optical transparency region. 

A comparison of the Fourier spectra (insets in Fig.~\ref{Figure1}b and \ref{Figure1}c) reveals that the frequency of the spin precession excited by pulses in resonance with the lattice mode is shifted as compared to the equilibrium value given by the ISRS driven reference. The sign of the shift $\Delta f$ depends on the initial magnetic configuration, being red in the AFM phase and blue in the WFM phases. To underscore the resonant character of the frequency shift we tune the photon energy of the pump pulse across the phonon resonance and extract the central frequency of the spin oscillations (Fig.~\ref{Figure1}d). The data acquired in both magnetic phases show that the onset of the frequency-shift follows closely the spectrum of the linear absorption of the B$_\textnormal{1u}$ phonon mode, revealing a correlation between light-driven lattice and spin dynamics. These non-linear dynamics of the magnetic soft mode contain important information on the effects of phonon-pumping on the magnetic potential. The red shift indicates a flattening of the potential energy in vicinity of the AFM minimum ($\phi=0^\circ$), which may lead to a phase instability. The blue shift observed in the WFM phase points at an increased curvature and phase stability in vicinity of the WFM minimum ($\phi=90^\circ$).

Time-resolved Fourier analysis indicates that the change in frequency occurs within a half-cycle of the first oscillation, defining an upper bound for the phonon-driven changes in the magnetic potential of about \SI{5}{\pico\second}. This value is nearly one order of magnitude smaller than the one reported for optical pumping of the $d$-$d$ electronic transitions~\cite{Yamaguchi2015}. Thus, the observed ultrafast response excludes heat-driven spin-lattice relaxation as the origin of the initial kinetics. Remarkably, the change in frequency persists for a time exceeding significantly the precession period, as shown in Fig.~\ref{Figure2}. We note that the relaxation time $\tau$ (inset Fig.~\ref{Figure2}) and its temperature dependence closely follow the characteristic equilibrium decay time of the spin precession (Fig.~S8).

\begin{figure*}[ht]
	\centering
	\includegraphics[width =\linewidth]{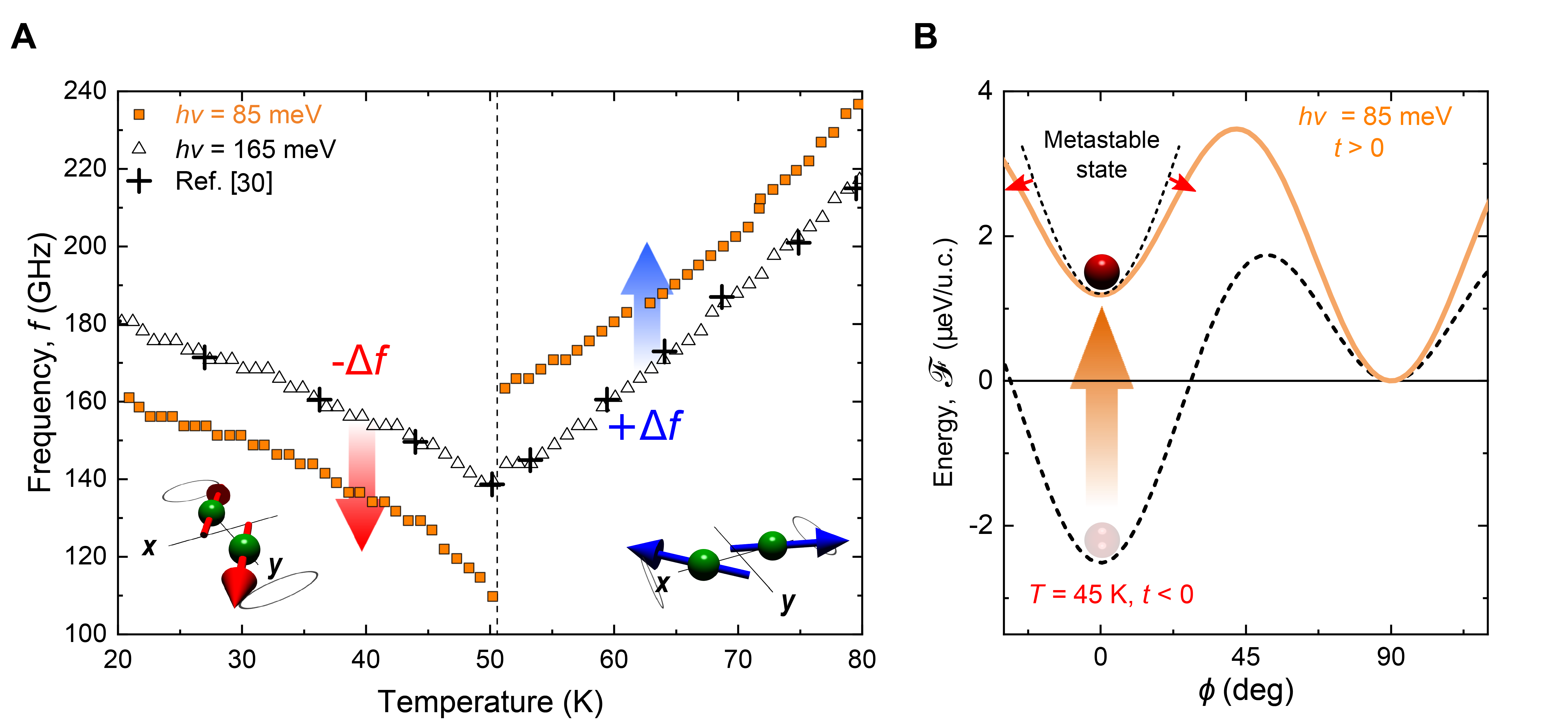}
	\caption{\textbf{Out-of-equilibrium metastable magnetic state.} (\textbf{A}) Frequency of the spin precession as a function of the temperature across the Morin phase transition for different photon energies of the pump excitation.
		%	after the light excitation with the photon energy tuned off-resonance (black triangular markers) and in-resonance (red and blue square markers) with the $B_{1u}$ phonon mode. The frequencies denoted by cross markers are data of sub-millimeter static spectroscopy from Ref.~\cite{Balbashov1985}. 
		%(\textbf{B}) Equilibrium magnetic potentials below (dashed) and above (solid) $T_\textnormal{M}$. 
		(\textbf{B}) Reconstructed magnetic potential before $t<0$ (dashed black) and after $t>0$ (solid yellow) the phonon-pumping. The position of the red ball represents the energy state of the system.
	} 
	\label{Figure3}
\end{figure*}

As shown in Fig.~\ref{Figure3}a, measurements of the frequency of the magnetic mode as a function of temperature reveal a cusp-like evolution in proximity to $T_\textnormal{M}$, in close agreement with Ref.~\cite{Balbashov1985}. In contrast, the frequency of the phonon-driven oscillations  at $T_\textnormal{M}$ is characterized by a pronounced discontinuity of more than \SI{50}{\giga\hertz} and frequencies well below the minimum value attainable at equilibrium.~%, underscoring the non-equilibrium nature of the light-induced state.
From the analysis of the temperature dependence of the phonon frequencies, we map out the magnetic potential~$\mathscr{F}$~(see supplementary text S4 for details) before ($t<0$) and after ($t>0$) laser excitation, see Figure~\ref{Figure3}b. Importantly, phonon-pumping significantly increases the energy of the AFM state simultaneously lowering the kinetic barrier, such that close to $T_\textnormal{M}$ the AFM phase may lose its stability.% We note that, in the AFM phase, phonon-excited DyFeO$_3$ exhibits magnon frequencies well below the minimum value attainable through a thermodynamic transformation, underscoring the non-equilibrium nature of the light-induced state. 

%This state is not accessible in the equilibrium because of the nucleation

%The fundamental difference is that phonon-pumping changes to the potential are so fast that system appears to be in the metastable overheated state AFM state.
%raising the temperature above $T_\textrm{M}$ in the equilibrium .  This is the origin of the new frequencies. 
%, while simultaneously reducing the barrier separating the AFM from the WFM phase.

%Note, that in the equilibrium changes in magnetic potential and phase stability,

\begin{figure*}[ht]
	\centering
	\includegraphics[width =\linewidth]{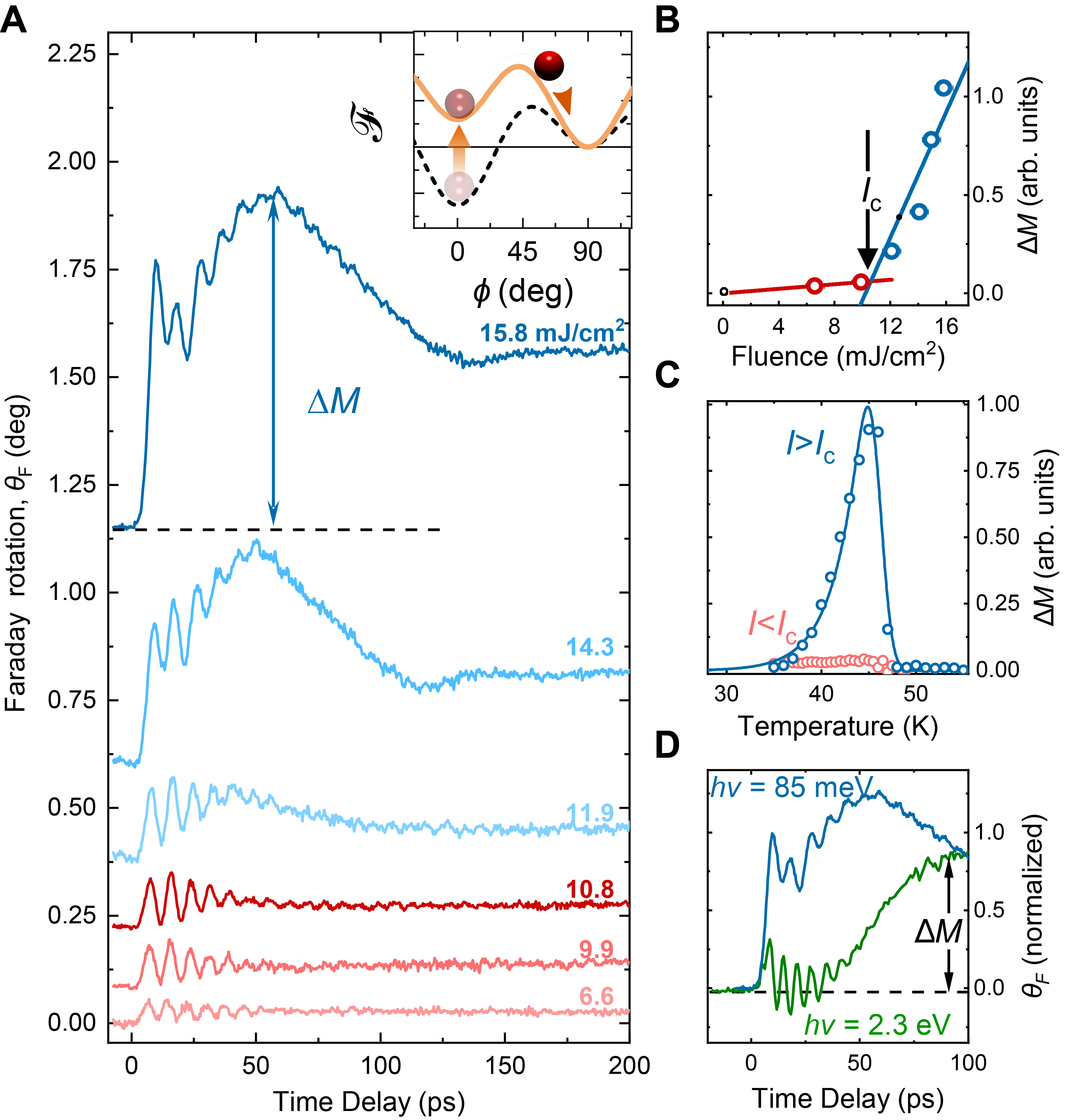}
	\caption{\textbf{Ultrafast phonon-induced magnetic phase transition.} (\textbf{A}) Time-resolved dynamics of the Faraday rotation $\theta_\textrm{F}$ close to $T_\textnormal{M}$ for various fluences of the pump. %The traces were obtained by measuring and subtracting the pump-induced Faraday rotation measured at applied magnetic fields $H = \pm$ \SI{0.4}{\tesla}. 
	Inset: Schematics of the ballistic reorientation. (\textbf{B}) Amplitude of the photo-induced magnetization $\Delta M$ as a function of the pump fluence. (\textbf{C}) $\Delta M$ as a function of the sample temperature for fluences above (blue curve) and below (red curve) the threshold $I_\textnormal{c}$. (\textbf{D}) A direct comparison of the dynamics of the magnetization initiated by a pump pulse with above-bandgap photon energy and in resonance with the phonon mode.}
	\label{Figure4}
\end{figure*}

The observation of light-induced metastability and the lowering of the kinetic barrier indicate that phonomagnetism can be used to promote an ultrafast spin-reorientation and drive a transformation from the AFM phase to the WFM phase (inset Fig.~\ref{Figure4}a). In order to explore this scenario we perform time-resolved measurements of the magnetization dynamics for a set of progressive pump fluences in vicinity of $T_\textnormal{M}$, as shown in Figure~\ref{Figure4}a. Two distinct regimes are observed above and below a critical fluence $I_\textnormal{c} \simeq $ \SI{10}{\milli\joule\per\centi\meter\squared} (see Fig.~\ref{Figure4}b). For pump fluences below $I_\textnormal{c}$ the pump drives a nearly harmonic response of spins mainly influencing its amplitude and frequency. However, upon exceeding the value of $I_\textnormal{c}$, the course of the magnetization dynamics changes dramatically. A long-lived (>\SI{250}{\pico\second}) offset, odd parity with respect to the sign of the applied field, develops within nearly a half-period of the spin precession(<\SI{5}{\pico\second}), marking the coherent emergence of a transient magnetization $\Delta M$. As the temperature is varied, it becomes apparent that the magnetisation peaks nearly at $T_\textnormal{M}$ and completely vanishes for $T > T_\textnormal{M}$ (see Fig.~\ref{Figure4}c). The presence of a well-defined fluence threshold, as well as the observation of the photo-induced macroscopic magnetization inherent to the WFM phase, are clear signatures that phonon pumping drives the coherent reorientation of spins accompanied by a change in the magnetic symmetry.

The phenomenological model indicates that all the observed dynamics are governed by phonon-induced changes in the strength of the \ce{Fe}-\ce{Dy} exchange (see supplementary information).~%, pointing at a specific microscopic mechanism of spin-lattice coupling 
We also use density functional theory calculations (see supplementary text S4) to model microscopic non-linear lattice dynamics and find that A$_g$ Raman distortions (a combination of antipolar Dy motions and oxygen octahedra rotations) have the strongest symmetry-allowed coupling with the B$_{1u}$ mode excited at resonance. Time-resolved measurements of the reflected probe pulse at twice the incident energy (second harmonic generation) are sensitive to structural dynamics (see supplementary text S4B) and are consistent with the non-linear excitation of these low-energy modes. Our calculations further indicate that a lattice distortion with the symmetry of the A$_g$ mode can strongly perturb the \ce{Fe}-\ce{Dy} exchange, underscoring a good agreement between the phenomenological and the microscopic model.

In Fig.~\ref{Figure4}d we compare the spin-reorientation transition obtained in DyFeO$_3$ via conventional photomagnetism (excitation with photon energy \SI{2.3}{\electronvolt}, above the material's band gap) with the phonomagnetism discussed here. We note that the photomagnetic response leads to the development of a net magnetisation on a time scale of 100 ps, consistent with a laser-induced heating of the electron, lattice and the rare-earth spin system \cite{Yamaguchi2015}. In contrast, the phonomagnetic route develops light-induced magnetism within a half cycle of the spin precession, below 5 ps. This time scale points at phonon-induced atomic displacements and reconfiguration of the anisotropic exchange interactions as the driving force of the spin alignment.

By optical pumping of lattice vibrations we have investigated a new route to reshape magnetic potentials and initiate ultrafast magnetic transitions. Using a phenomenological description and first-principles calculations we have demonstrated that the mechanism is based on a lattice-driven change of the magnetic anisotropy (exchange interaction between rare-earth and transition metal ions). We anticipate that a similar mechanism will be active in other magnetic materials featuring anisotropic interactions such as rare-earth based compounds (manganites, vanadates and orthochromites) and type-\textnormal{II} multiferroics ~\cite{Cheong2007,Tokura2014}. Our findings highlight new avenues for the manipulation of ferroic order on ultrafast timescales and provide a new platform for magnonics.

\textbf{Acknowledgments:}
We thank S.~Heirman for assistance in measuring linear absorption in the mid-infrared spectral range and T.~C.~van~Thiel for fruitful discussions.

\textbf{Funding:} This work was supported by the EU through the European
Research Council, Grant No. 677458 (AlterMateria),
The Netherlands Organization for Scientific Research
(NWO/OCW) as part of the Frontiers of Nanoscience
program (NanoFront), and VENI-VIDI-VICI program,
the European Union's Seventh Framework Program (FP7/2007-2013)/ERC Grant Agreement No. 339813
(Exchange), the program Leading Scientist of the Russian
Ministry of Science and Higher Education
(14.Z50.31.0034), the Ministry of Science and Higher Education of the Russian
Federation in the framework of Increase Competitiveness
Program of NUST MISiS (No. K2-2017-005), implemented
by a governmental decree dated 16th of March 2013, N211,
and by the National Academy of Sciences of Ukraine via
Project No. 1/17-N. E.B. and A.S. thank the FRS-FNRS, ARC AIMED project, the C\'{E}CI supercomputer facilities (Grant No. 2.5020.1) and Tier-1 supercomputer of the F\'ed\'eration Wallonie-Bruxelles funded by the Walloon Region (Grant No. 1117545).

\textbf{Author contributions:} A.D.C. conceived the project with D.A. D.A and J.R.H. carried out the experiments and analyzed the data. D.A., R.V.M. and A.V.K. identified the material system for the project and contributed to the analysis. B.A.I and Y.B. contributed to theoretical treatment of experimental results. A.S. and E.B. performed the DFT calculations. All authors discussed the results. The manuscript was written by D.A., J.R.H. and A.D.C. with feedback from all coauthors.

\textbf{ Competing interests:} The
authors declare no competing interests. Data and materials
are available on a reasonable request.

\FloatBarrier 

%%%%%%%%%%%%%%%%%%%%%%%%%%%%%%%%%%%%%%%%%%%%%%%%%%%%%%%%%%%%%%%%%%%%%
%%%%%%%%%%%%%%%%%%%%% Supplementary Information %%%%%%%%%%%%%%%%%%%%%
%%%%%%%%%%%%%%%%%%%%%%%%%%%%%%%%%%%%%%%%%%%%%%%%%%%%%%%%%%%%%%%%%%%%%
\pagebreak
\widetext
\setcounter{section}{0}
\setcounter{equation}{0}
\setcounter{figure}{0}
\renewcommand{\thefigure}{S\arabic{figure}}
\renewcommand{\theequation}{S\arabic{equation}}
\renewcommand{\thetable}{S\arabic{table}}

\section*{Materials and Methods}
\section*{S1. Experimental setup}

\begin{figure}[H]
	\centering
	\includegraphics[width =0.75\linewidth]{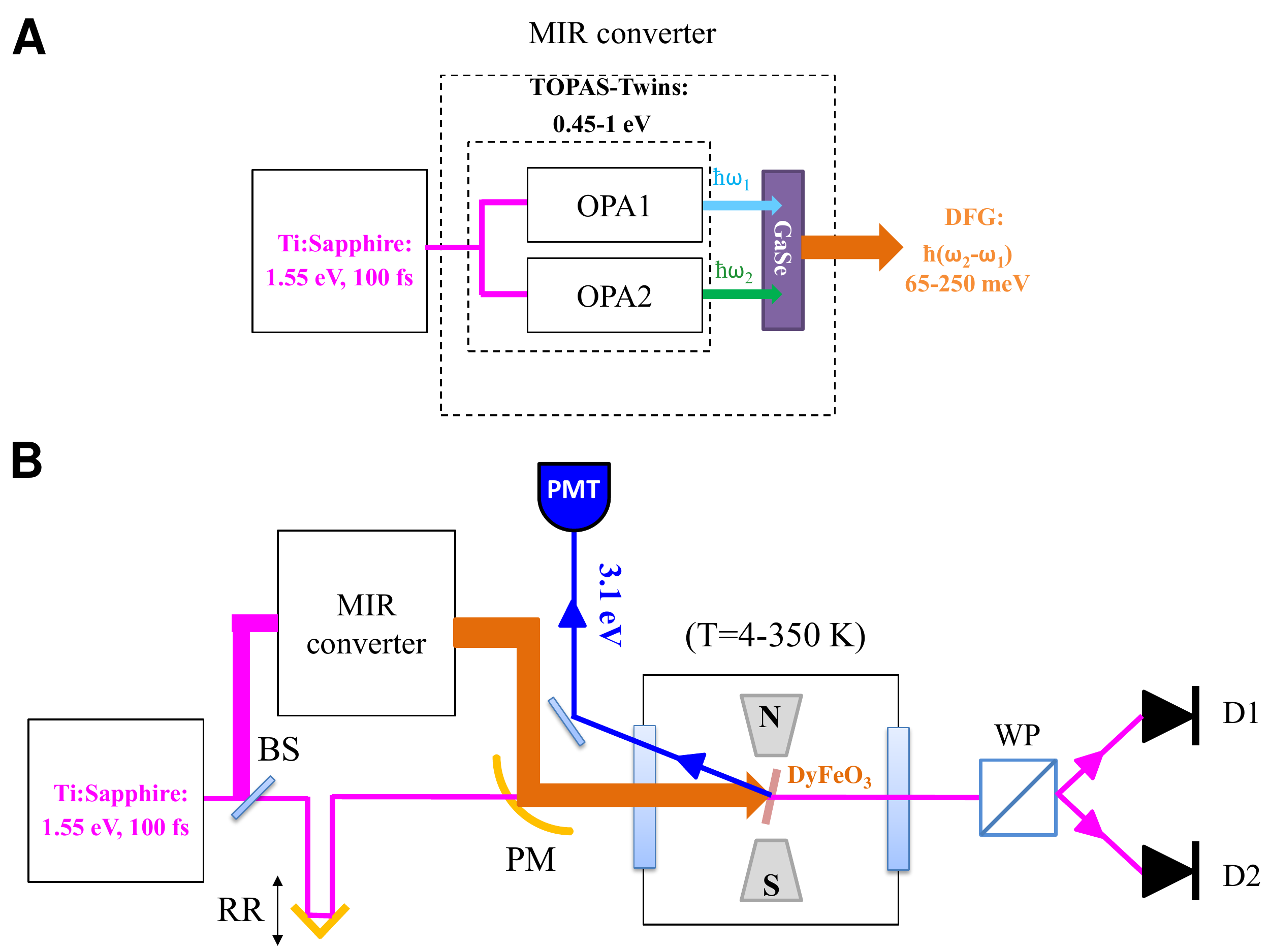}
	\caption{\textbf{Experimental setup} (\textbf{A}) Schematic overview of the experimental setup (MIR converter) used to generate intense MIR transients from the near-infrared ultrashort light pulses. (\textbf{B}) A schematic representation of the experimental setup described in the methods section of the main text. BS-beam splitter, RR-gold retroreflector mounted on a motorized precision delay stage, PM-parabolic mirror, WP-Wollaston prism, D1,2-silicon photodetector, N and S define the poles of the resistive magnet placed inside the cryostation, PMT is a photomultiplier tube used for detection of the second harmonic signal at double the probe photon energy.} 
	\label{FigureS1}
\end{figure}

The intense pump pulses with wavelengths in the mid-infrared (MIR) spectral range used in this experiment, were generated via difference frequency generation (DFG) in a \ce{GaSe} crystal using the output beams of two commercially available, independently tunable optical parametric amplifiers (OPAs) integrated into a single housing (Light Conversion, TOPAS-Twins). The OPAs were pumped by a commercially available amplified Ti:Sapphire laser system (Coherent, Astrella) delivering pulses at a \SI{1}{\kilo\hertz} repetition rate with a duration of \SI{100}{\femto\second} and a photon energy of \SI{1.5}{\electronvolt}. The OPAs were seeded by the same white light generated in a sapphire crystal, which ensures separately tunable, but phase-locked output pulses \cite{Sell2008} with photon energies in the range of \SIrange[range-units = single, range-phrase = {~to~}]{0.45}{1}{\electronvolt}. As a result, when the pulses are mixed in the GaSe crystal, the generated MIR pulses are carrier envelope phase (CEP) stable \cite{Baltuska2002}, linearly polarized and their energy lies in the range from \SIrange[range-units = single, range-phrase = {~to~}]{65}{250}{\milli\electronvolt} with an average pulse duration of around \SI{200}{\femto\second}. In the experiments, the mid-infrared pulses were focused onto the sample surface to a spot with a diameter of about \SI{150}{\micro\meter}, using an off-axis parabolic mirror. The scheme of the setup for generation of the MIR pulses (MIR converter) is shown in Fig.~\ref{FigureS1}A. The experimental geometry employed for measuring pump-induced dynamics is shown in Fig.~\ref{FigureS1}B. The polarization rotation $\theta_\textnormal{R}$ of the linearly polarized probe pulse, imprinted by the magneto-optical Faraday effect, was measured with the help of a polarization bridge (Wollaston prism) and a balanced photodetector. The pump-induced changes to the crystal symmetry of \ce{DyFeO3} were monitored via time-resolved measurements of the reflected probe pulse at twice the incident energy (\SI{3.1}{\electronvolt}), detected in the reflection geometry.

\FloatBarrier

\newpage
\section*{Supplementary Text}

\section*{S2 First-order spin reorientation phase transition of the Morin type in $\text{DyFeO}_3$}
\setcounter{subsection}{0}

\subsection{Phase diagram of the phase transition}

Below the N\'eel temperature $T_\textnormal{N}=\SI{650}{\kelvin}$, the \ce{Fe^{3+}} spins in \ce{DyFeO3} are ordered antiferromagnetically. The magnetic phase diagram of \ce{DyFeO3} includes a collinear antiferromagnetic (AFM) phase with the spins oriented along the $y$-axis and a weakly ferromagnetic (WFM) phase with the spins oriented along the $x$-axis \cite{Treves1965}. The AFM phase is stabilized at low temperatures ($T<T_\textnormal{M}$ = \SI{51}{\kelvin}) and can be transformed to the WFM phase either by an increase in temperature or by application of an external magnetic field applied along the $z$-crystallographic axis.\cite{Afanasiev2015} The transition between the phases proceeds as a first-order phase transition. The measurements of the static Faraday rotation as a function of the applied magnetic field $B$ for various temperatures $T$ shown in Fig.~\ref{Fig:EquiPotential}A reveal the phase diagram of the magnetic states.

The phase transition demonstrates a very narrow (<\SI{0.2}{\kelvin}) temperature hysteresis~\cite{Belov1976,Eremenko1987,Maziewski1977,Gnatchenko1989,Szymczak1983,Gnatchenko1981}. In Refs.~\cite{Eremenko1987,Szymczak1983} it was assumed that this nearly hysteresis-free character of the transition arises from a high sensitivity of the spin-orientation phase transition to lattice deformations and imperfections of the crystals. These factors contribute to prompt nucleation of the new magnetic phase, once the energy balance is altered, making the hysteresis extremely narrow in vicinity of $T_\textnormal{M}$.

\subsection{Free energy potential of the phase transition}

The spontaneous spin-reorientation transition in \ce{DyFeO3} in zero magnetic field is governed by the magneto-crystalline energy:

\begin{equation}
F\left(\phi,T\right)=\frac{1}{2}K_2(T)\cos^2\phi-\frac{1}{4}K_4\cos^4\phi,
\label{Eq:potential}
\end{equation}
which depends on the temperature $T$ and the angle $\phi$ that the spins form with the $y$-axis~\cite{Zvezdin1979,Balbashov1985}. The magnetic anisotropy parameters $K_2$ and $K_4$ originate from the temperature-dependent $K_2$ exchange interaction between the iron and dysprosium ions and the temperature-independent magnetic anisotropy of the iron ion $K_4$. 

\begin{figure}
	\centering
	\includegraphics[width =1.0\linewidth]{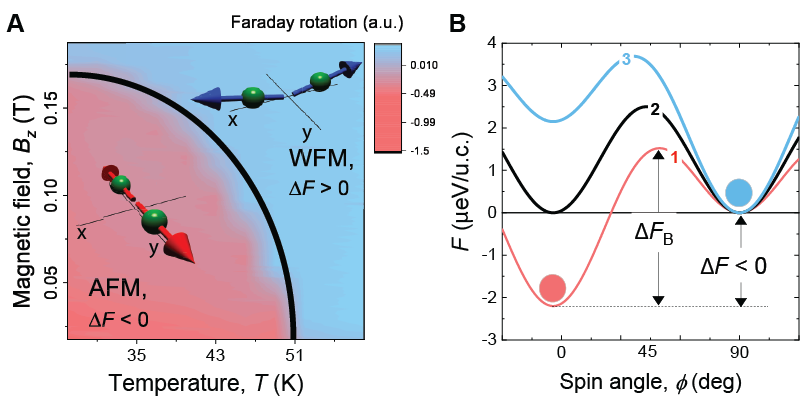}
	\caption{\textbf{Spin-reorientation transition of Morin type in \ce{DyFeO3}.} (\textbf{A}) Phase diagram of the magnetic states of \ce{DyFeO3} as a function of the magnetic field $B$ applied along the $z$-axis and temperature $T$ as measured using the static magneto-optical Faraday effect. (\textbf{B}) The energy profile of the magnetic potential vs. angle $\phi$ for three different ratios of the constants $K_2$ and $K_4$, corresponding to the stable AFM state: $\frac{K_2}{K_4}=0.39$ (1, $T<T_\textnormal{M}$), the Morin point: $\frac{K_2}{K_4}=0.5$ (2, $T = T_\textnormal{M}$), and the stable WFM state: $\frac{K_2}{K_4}=0.61$ (3, $T > T_\textnormal{M}$) phase in equilibrium. The red and blue ball depict the system's ground states.
	} 
	\label{Fig:EquiPotential}
\end{figure}

Figure~\ref{Fig:EquiPotential}b shows that the relative energy of the AFM and WFM phases $\Delta F$ ($\Delta F<0$ when the energy of the AFM phase is lower than that of the WFM) as well as the height of the kinetic potential barrier $\Delta F_\textnormal{B}$ is controlled by the ratio of the two positively defined phenomenological parameters $K_2$ and $K_4$. We note that this potential cannot account for the hysteresis-free character of the phase transition and rather represents the free-energy potential of a single nucleus. The adiabatic phase changes in the equilibrium are governed by the $\Delta F$ value and not by the height of the kinetic barrier $\Delta F_\textnormal{B}$. 

\subsection{Relation between the curvature of the magnetic potential and frequency of the soft mode spin precession.}

\begin{figure}[H]
	\centering
	\includegraphics[width=1.0\linewidth]{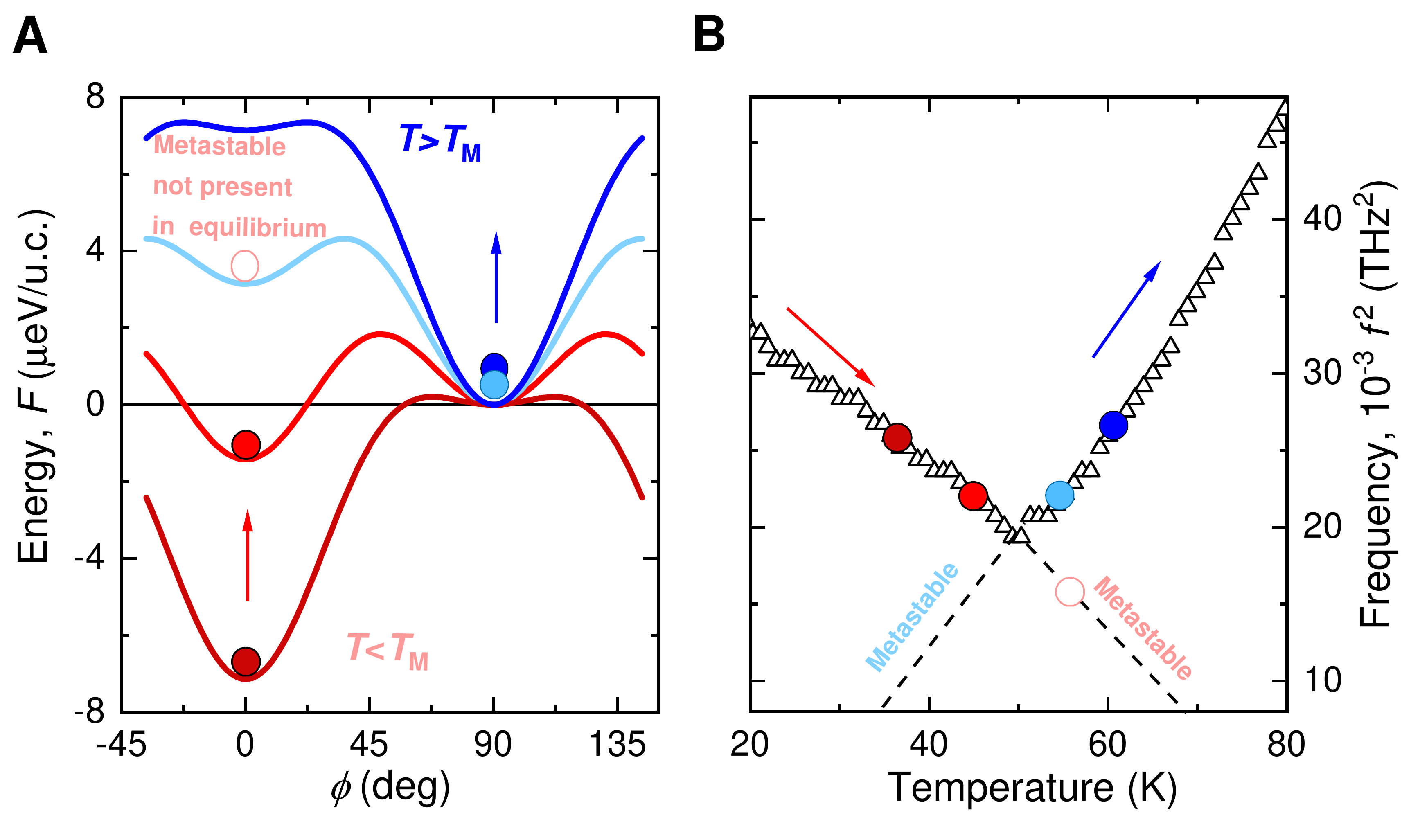}
	\caption{\textbf{Correspondence between the shape of the potential and frequency of the soft mode.} (\textbf{A}) Evolution of the magnetic potential with temperature (direction from low to high temperatures are denoted by the arrows) across the Morin temperature. The position of the solid ball indicates the system's ground state. The position of the open ball indicates the system's metastable state that is not present in equilibrium due to dominance of the nucleation processes. (\textbf{B}) Squared frequency of the soft mode precession as a function of temperature. The dashed lines denote frequencies of the spin precession in the metastable state which are not realized in the equilibrium. The balls indicating the ground state in (A) are mapped on the temperature dependence to match the frequency with the potential and the ground state.
	} 
	\label{Fig:freqeq}
\end{figure}

Using the equilibrium magnetic potential $F(\phi,T)$ (Eq.~\ref{Eq:potential}), the frequency of the antiferromagnetic spin precession $f$, the soft mode of the phase transition, in the AFM ($T<T_\textnormal{M}$, $\phi_0=0^{\circ}$) and the WFM ($T>T_\textnormal{M}$, $\phi_0=90^{\circ}$) phases can be obtained from:
\begin{equation}
\begin{aligned}
\frac{\omega^2}{\omega_\textnormal{exc}}=\frac{\gamma}{M_0}\frac{\partial^2 F}{\partial \phi^2}\bigg\rvert_{\phi=\phi_0}=
\frac{\gamma}{M_0}\cdot
\begin{cases}
-K_2+K_4, &\;\phi_0=0^\circ \\
K_2,&\;\phi_0=90^\circ
\end{cases}.
\end{aligned}
\label{Eq:Frequency}
\end{equation}
Here, $\omega=2\pi f$, with $\omega$ and $f$ the angular and regular spin precession frequency respectively. $\omega_\textnormal{exc}$ is the so-called exchange frequency, defined as the angular frequency of the spin precession in the exchange field $H_\textnormal{exc}$ of the magnet, such that $\omega_\textnormal{exc}=\gamma H_\textnormal{exc}$, where $\gamma$ is the gyromagnetic ratio. $M_0 = \frac{5}{2} g \mu_\textnormal{B} /  \frac{V_{u.c.}}{4}$ is the saturation magnetization of a single Fe$^{3+}$ sublattice, given by the $g$-factor and Bohr magneton $\mu_\textnormal{B}$ divided by the unit cell volume $V_{u.c.}$ containing 4 iron ions. Comparing the potentials depicted in Fig.~\ref{Fig:freqeq}A, one can see that upon transition from the AFM to WFM phases softening of the frequency in vicinity of the $T_\textnormal{M}$ (Fig.~\ref{Fig:freqeq}B) is caused by flattening of the potential function at the AFM minimum. We note that complete flattening, manifested by the frequency going to zero, does not occur in equilibrium, see the dashed line in Fig.~\ref{Fig:freqeq}B. This is because  as the temperature is increased to $ T > T_\textnormal{M}$ the nucleation process transfers all the phase to the energetically more favorable WFM minimum ($\phi_0=90^\circ$). The similar situation occurs with the WFM minimum when $T_\textnormal{M}$ is approached from above. 

\newpage

\section*{S3. Light-induced excitation of coherent spin precession}
\setcounter{subsection}{0}

\subsection{Crystal structural and absorption spectrum of $\text{DyFeO}_3$}

\begin{figure*}[h]
	\centering
	\includegraphics[width =0.8\linewidth]{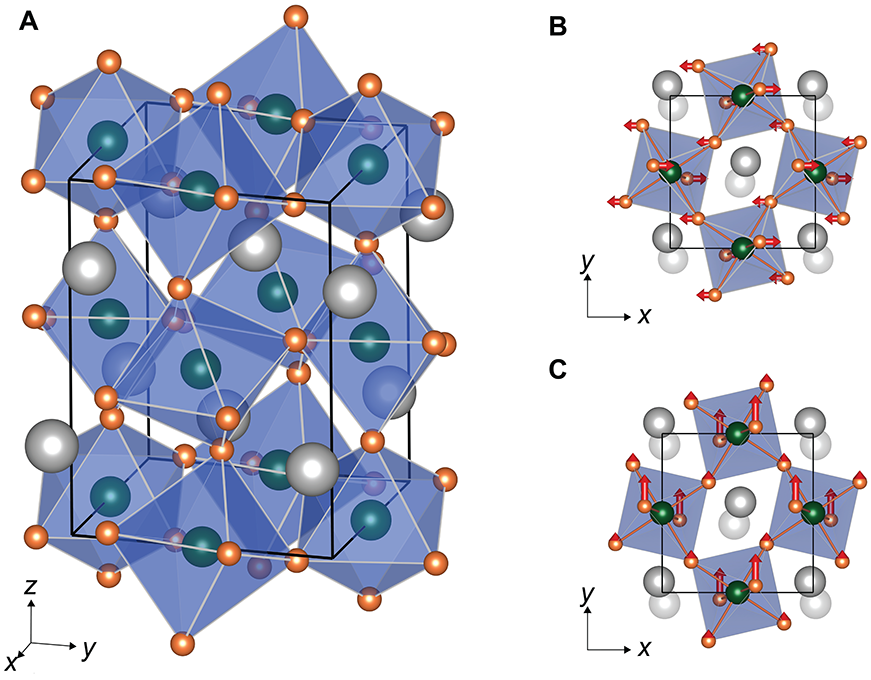}
	\caption{\textbf{Structure and optical phonon modes.} (\textbf{A}) Crystal structure of \ce{DyFeO3}. Grey/green/orange spheres represent dysprosium/iron/oxygen atoms. (\textbf{B} and \textbf{C}) Polar component of the eigenvectors of the highest-frequency infrared-active $B_\textnormal{1u}$ (B) and $B_\textnormal{3u}$ (C) phonon modes. } 
	\label{Fig:DFO_structure}
\end{figure*}

\ce{DyFeO3} crystallizes in a perovskite orthorhombic structure (\textit{Pnma} space group). Its crystal structure is shown in \ref{Fig:DFO_structure}a. In our experiments, the ultrashort MIR pulses were tuned in resonance with the highest-frequency, nearly-degenerate $B_\textnormal{1u}$ and $B_\textnormal{3u}$ phonon modes. The atomic motion comprising these phonon modes are shown in \ref{Fig:DFO_structure}b,c. The electric field of the ultrashort pulses (travelling in the $z$-direction) can couple to these modes, which have their dipole moment in the $x$-$y$ plane. 

\begin{figure*}[h]
	\centering
	\includegraphics[width =0.8\linewidth]{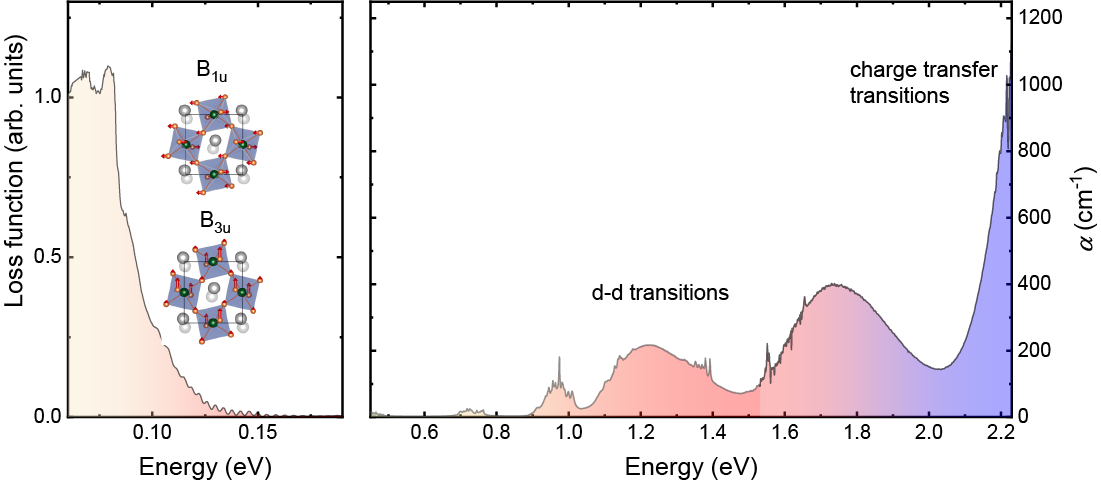}
	\caption{\textbf{Optical absorption spectrum of \ce{DyFeO3}.} Loss function and optical absorption for the \SI{60}{\micro\meter} thick (001) \ce{DyFeO3} sample in a broad energy range. The B$_\textnormal{1u}$ and B$_\textnormal{3u}$ phonon modes are nearly degenerate and dominate the loss function at energies below \SI{0.15}{\electronvolt}} 
	\label{Fig:DFO_absorption}
\end{figure*}

In order to understand and compare the effect of excitation with pump pulses with a different photon energy, we studied the steady-state absorption spectrum of \ce{DyFeO3}. For the infrared wavelengths, the total transmission of the sample was measured using a Nicolet 6700 Fourier transform infrared (FTIR) spectrometer, which led to the loss function in Fig.~\ref{Fig:DFO_absorption} (left panel). At higher energies, the absorption of the sample shown in the right panel of Figure~ \ref{Fig:DFO_absorption} was determined using a white light source and a set of Oceanview spectrometers. The broad absorption peaks below \SI{2.0}{\electronvolt} are due to \ce{Fe^{3+}} $d$-$d$ transitions, while the onset of the large peak at high energies is due to charge-transfer excitations.~\cite{Wood1970}

\subsection{Dependence on photon energy of the pump pulse}

\begin{figure*}[h]
	\centering
	\includegraphics[width =0.8\linewidth]{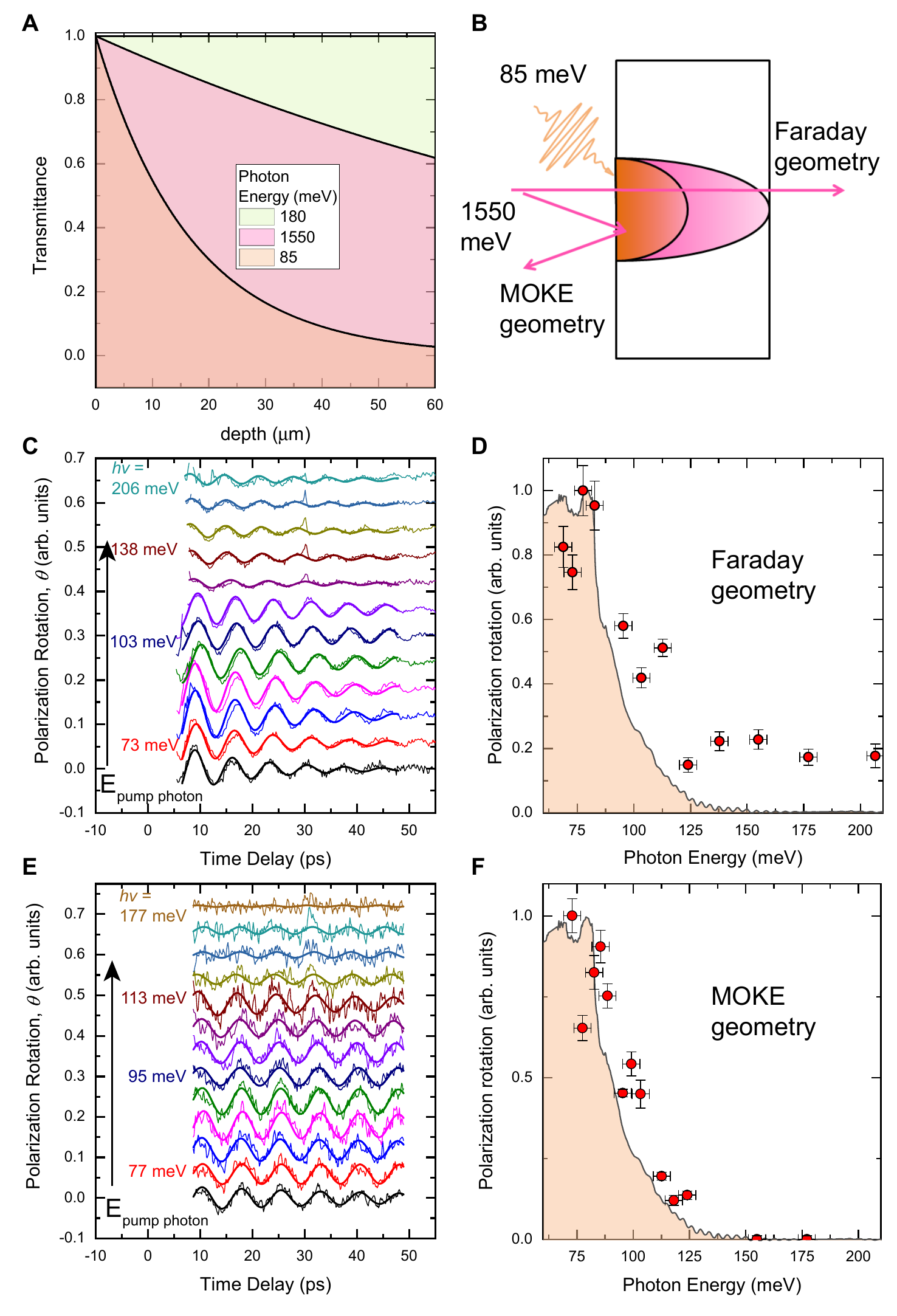}
	\caption{\textbf{Pump photon energy dependence} (\textbf{A}) Optical absorption profile for the photon energy corresponding to the probe (\SI{1550}{\milli\electronvolt}) and two energies of the pump pulse. (\textbf{B}) Schematic of the penetration depth of the pump pulse (\SI{85}{\milli\electronvolt}) and the probe pulse (\SI{1550}{\milli\electronvolt}) and the two measurement geometries. (\textbf{C} and \textbf{E}) Transient polarization rotation (showing spin dynamics) of the \SI{1550}{\milli\electronvolt} probe pulse after excitation with pump pulses of different energy in the transmission (C) and reflection (E) geometry. Thick solid lines are damped sine fits. (\textbf{D} and \textbf{F}) Amplitude of the pump-induced spin precession measured in the transmission geometry via the Faraday rotation (D) and the reflection geometry via the magneto-optical Kerr effect (F) as a function of the pump photon energy. Amplitudes were extracted from the fits of the data in (C) and (E). The light-brown shaded area corresponds to the loss function of the sample, as introduced in Fig.~1D (main text).} 
	\label{Fig:S2}
\end{figure*}

To study effect of the pump photon energy on the excitation of the coherent spin precession, we tuned the photon energy of the pump pulse across the phonon resonance and extracted the amplitude of the spin oscillations, see Fig.~\ref{Fig:S2}.
The volume of the sample which is probed or excited is defined by the penetration depth of the incident light. This quantity is strongly dependent on the photon energy. The dependency is especially pronounced in vicinity of optical resonances such as formed by the absorption of the infrared-active phonon mode. At these energies, the penetration depth of the incident light is significantly decreased. In our experiment the photon energy of the probe light remained constant throughout all measurements, implying that the probed volume is also constant. The pump energy, on the other hand, was varied in the broad energy range including the phonon absorption range. As a consequence the penetration depth of the pump, and therefore the excited volume, changed strongly (see Fig.~\ref{Fig:S2}A). 

To avoid this mismatch, we performed a set of the experiments in a reflection geometry employing the polar magneto-optical Kerr effect (p-MOKE), see Fig.~\ref{Fig:S2}B. In reflection only a nearly homogeneously excited thin top layer of the sample is probed. Hence these measurements form a way to determine the genuine magnetization dynamics as a function of the pump photon energy. 

Figure~\ref{Fig:S2}c,d show the pump-induced spin precession and extracted amplitude as a function of the pump photon energy measured at constant power of the pump beam as performed in the Faraday transmission geometry. Figure~\ref{Fig:S2}e,f show the results of a similar measurement, in this case performed in the reflection (MOKE) geometry.

A significantly more pronounced resonant character of the excited spin precession as a function of the photon energy is seen for the MOKE experiment compared to the Faraday rotation. We note that in the MOKE experiment no excitation of the spin precession was observed after off-resonant excitation. In contrast, in the Faraday experiment the spin oscillations were also clearly observed for the off-resonant excitation, which reveals that their presence in the Faraday signal is a result of the integration of the magneto-optical signal across the sample thickness. To summarize, the reflection geometry reveals the highly resonant character of the excitation of the spin precession as function of the pump photon energy.

\FloatBarrier

\subsection{Dependence on the orientation of the pump polarization plane}

\begin{figure}[H]
	\centering
	\includegraphics[width =\linewidth]{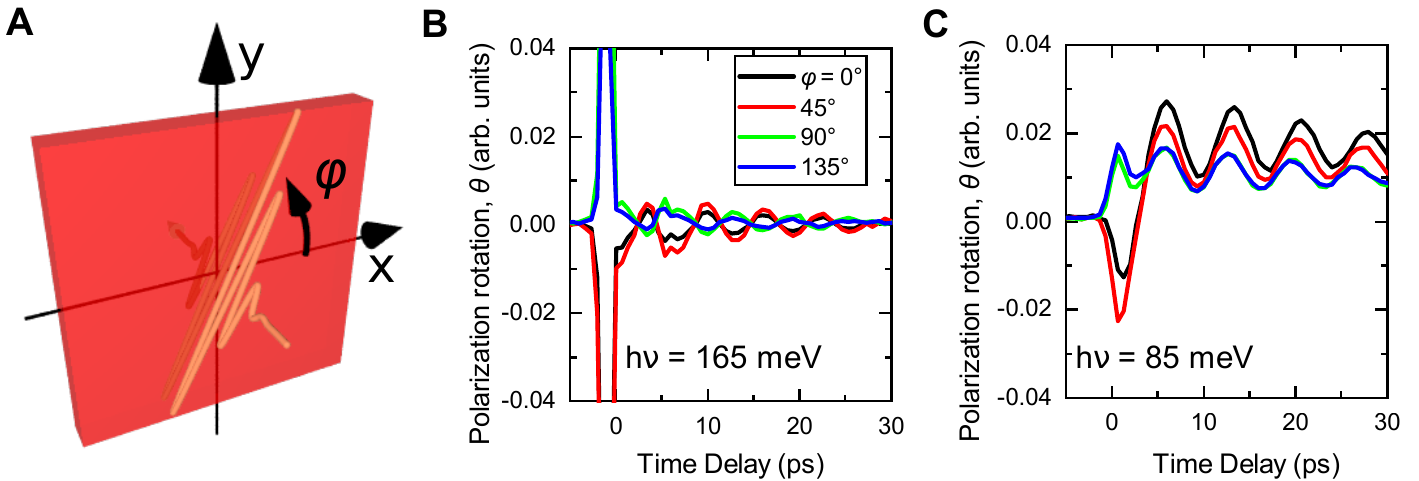}
	\caption{\textbf{Pump polarization dependence.} (\textbf{A}) The $z$-cut \ce{DyFeO3} with in-plane $x$ and $y$ crystal axes. The polarization plane angle $\phi$ of the pump pulse with respect to the $y$-axis can be varied. (\textbf{B} and \textbf{C}) Pump-induced, time-resolved Faraday rotation at various orientations of the polarization plane of the pump pulse. The data is presented for the off-resonant excitation at the pump photon energy of \SI{165}{\milli\electronvolt} (B) and for the resonant pumping at the pump photon energy of \SI{85}{\milli\electronvolt} (C). The sample was in the AFM phase, with a temperature of \SI{30}{\kelvin} and an applied magnetic field of \SI{0.4}{\tesla}. The strong polarization-dependent peak in vicinity of $t$=0 is due to the ultrafast Kerr effect.} 
	\label{Fig:S3}
\end{figure}

To reveal the excitation mechanism of the spin precession at different energies of the pump photon, we focus on pump pulses having photon energies either resonant or off-resonant with the absorption line of the phonon mode. We studied the dependence of the spin precession on the polarization of the pump pulse for off- (\SI{165}{\milli\electronvolt}) and on-resonant pumping (\SI{85}{\milli\electronvolt}). Measurements were performed for a set of four distinct orientations of the polarizations of the pump pulse. The orientations correspond to the electric field of the pump pulse oriented either along or in-between the sample's in-plane crystallographic $x$- and $y$-axis (see Fig.~\ref{Fig:S3}A) 

In the case of off-resonant pumping, we find that the orientation of the polarization of the pump pulse determines the initial phase of the spin precession phase, as shown in Fig~\ref{Fig:S3}b. The polarization dependence is reminiscent of the one expected for the inverse Cotton-Mouton effect (ICME)~\cite{Kalashnikova2007,Afanasiev2016}, a type of impulsive stimulated Raman scattering (ISRS), via which coherent spin precession is excited by linearly polarized ultrashort pulses. 
%According to this mechanism, the electric field of the pump pulse effectively acts as an instantaneous magnetic anisotropy, and its selection rules are determined by the material's magneto-optical constants and the properties of the spin precession itself. Hence, the pulse excites no spin precession if its polarization direction coincides with the magnetic order parameter oriented along the $y$- and $x$-directions in the AFM and WFM phases, respectively. At the same time, pulses with polarizations oriented at $\pm$ \ang{45} with respect to the crystal axes excite spin precession with opposite initial phases. We observe a flip of the precession phase upon change of the pump polarization from +\ang{45} (green) to -\ang{45} (black) in agreement with the ICME mechanism. Spin precession and a similar phase flip are also observed after excitation with the pump polarization oriented along the sample axes and can be explained by the strong optical birefringent properties of the rare-earth orthoferrites. The birefringence affects the initial polarization of the pump pulse during propagation in the sample.
In striking contrast to off-resonant pumping, no effect of the pump polarization on the phase of the spin precession is observed when the sample is pumped at the photon energy matching the phonon resonance, as is shown in Fig.~\ref{Fig:S3}C. A slight difference in frequency of the observed oscillations is explained by a difference in fluence of the pump pulse used to excite the spin dynamics (\SIlist[list-units = single]{9.2;11}{\milli\joule\per\centi\meter\squared}). The absence of any dependence on the pump polarization is a strong indication that a different excitation mechanism is at play.

%\section{Highly anharmonic magnetic response for the above threshold fluences of the resonant pumping }
%
%\begin{figure*}
	%\centering
	%\includegraphics[width =\linewidth]{S4FourierPT.eps}
	%\caption{\textbf{Fourier analyses of the phase transition} FFT spectra of the magnetization dynamics shown in Fig.~3a The numbers depict different bands. Central frequency of the 1$^\text{st}$ band as a function of the pump fluence. Panels (c), (d), (e) show spectral weight of 1$^\text{st}$, 2$^\text{nd}$ and 3$^\text{rd}$ bands as a function of the pump fluence. } 
	%\label{FigureS4}
%\end{figure*}
%
%In the main text we reported observation of the light-induced weakly ferromagnetic state for the pump fluence above threshold value ($F>F_c$), see Fig.~3a. Here, we show analysis of the corresponding Fourier spectra for fluences of the pump above and below the threshold $F_c$. The FFT reveals presence of three distinct bands, , see Fig.~\ref{FigureS4}. The first band can be attributed to the spin precession and demonstrates red-shift upon the fluence increase. Interestingly, the spectral weight integrated over this bands demonstrates highly nonlinear dependence on the pump fluence. The second band is centered at around 30 GHz and has bandwidth of around 30 GHz. We attribute this band to relaxation of the light-induced anisotropy which has lifetime on the order of 30 ps. The third situated at the lowest frequency corresponds to the emergence of the magnetization inherent to the WFM state and .
%

%\blindtext
\subsection{Temperature dependence}

\begin{figure}
	\centering
	\includegraphics[width =0.9\linewidth]{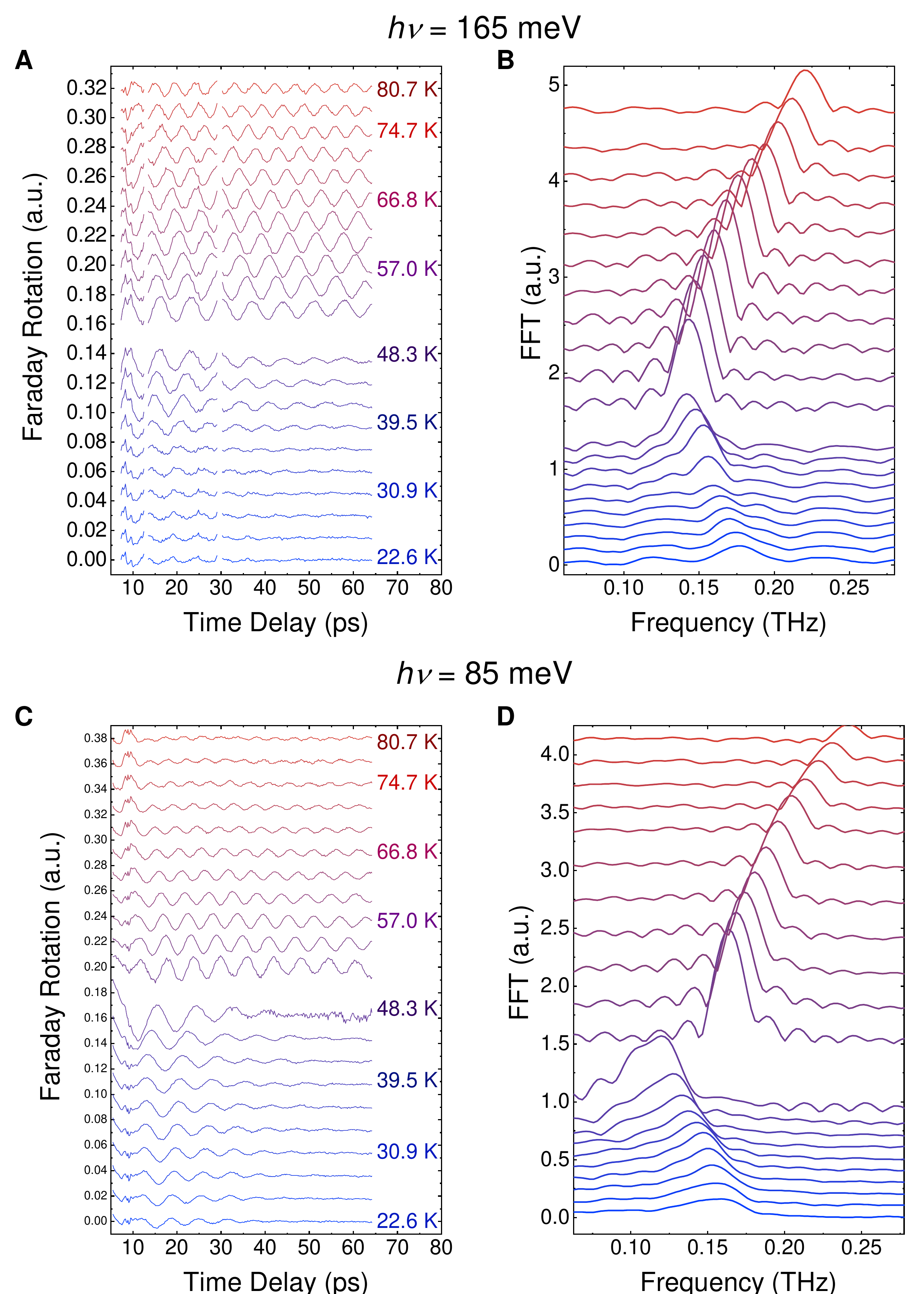}
	\caption{\textbf{Spin precession for various temperatures.} (\textbf{A} and \textbf{C}) The oscillatory part of the probe polarization rotation as a function of delay time between the pump excitation \SI{165}{\milli\electronvolt} (A), \SI{85}{\milli\electronvolt} (C) and the probe at different sample temperatures. (\textbf{B} and \textbf{D}) The corresponding Fourier spectra of the time-domain dynamics shown in panels (A), (C).   
	} 
	\label{Fig:S10_SpinFreq}
	
\end{figure}

In the main text (Figure 3a) we reported the spin precession frequency obtained after exciting the \ce{DyFeO3} sample at different pump photon energies as a function of temperature. Here we report the time-domain dynamics of this mode following excitation energy in resonance with the phonon (\SI{85}{\milli\electronvolt}) and off-resonance in the transparency range (\SI{165}{\milli\electronvolt}) from which the frequencies were extracted by taking the spectra of oscillatory signals. Both the time-domain dynamics and the Fourier spectra are shown in Fig.~\ref{Fig:S10_SpinFreq}. The Fourier spectra show clear maxima corresponding to the spin precession mode, with a softening close to the phase transition temperature around \SI{51}{\kelvin}. 

\FloatBarrier 

\subsection{Dependence on the fluence of the pump pulse}

\begin{figure}[h]
	\centering
	\includegraphics[width =\linewidth]{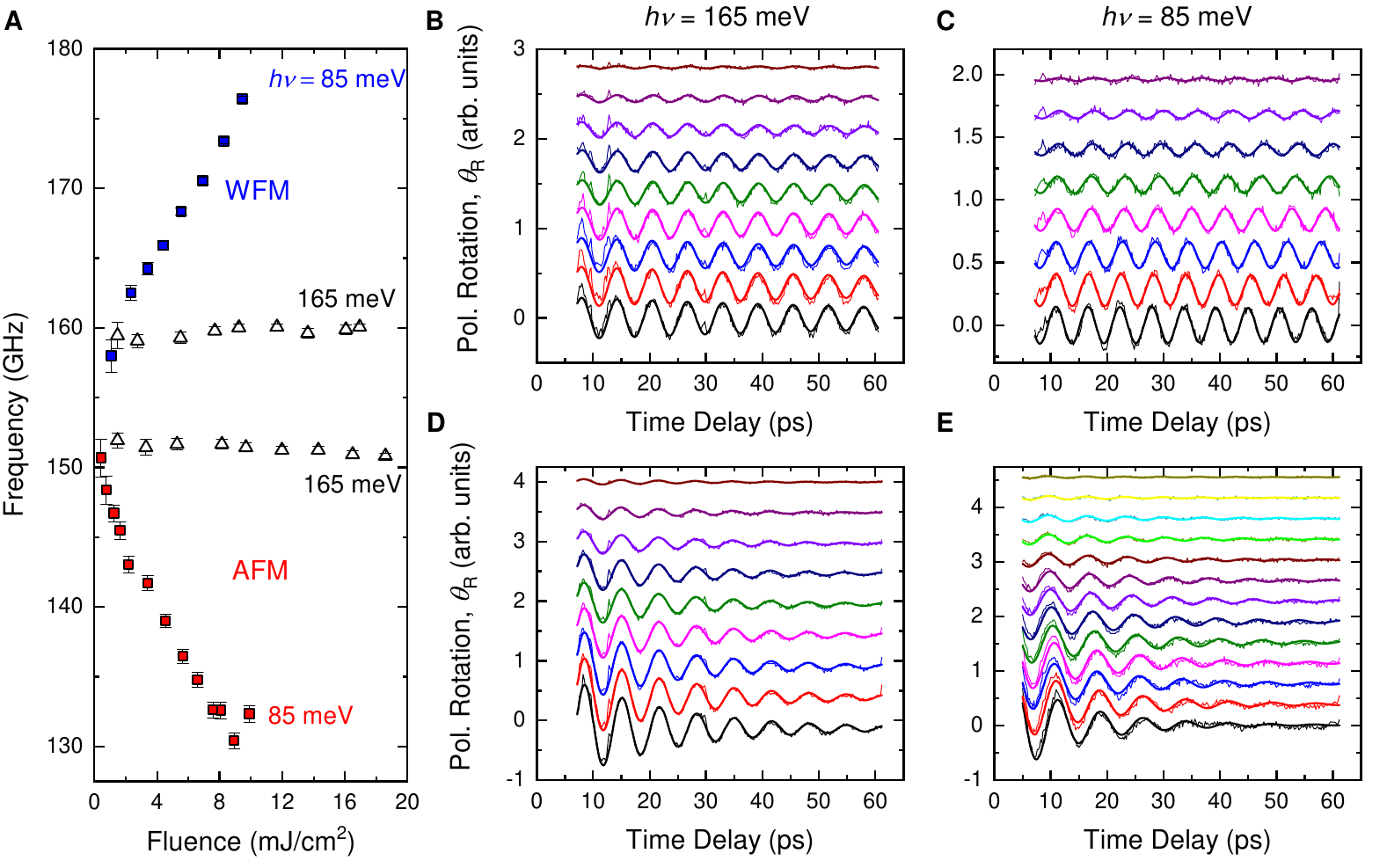}
	\caption{\textbf{Fluence dependence of the spin precession frequency.} (\textbf{A}) Frequency of the spin oscillation as function of the laser fluence. The open and solid points correspond to measurements at a pump photon energy of \SI{165}{\milli\electronvolt} and \SI{85}{\milli\electronvolt} respectively. The red and blue data points correspond to the sample being in the AFM and WFM phases, respectively. (\textbf{C}-\textbf{E}) Transient polarization rotation in the WFM phase after excitation with \SI{165}{\milli\electronvolt} (B) and \SI{85}{\milli\electronvolt} (C) pump pulses showing clear spin precession. The same measurements for these pump pulses in the AFM phase are shown in (D) and (E) respectively. The frequencies of the solid (exponentially damped) sine fits are shown in panel (A).
	}
	\label{Fig:FluenceFreq}
\end{figure}

The spin precession frequencies were extracted from the waveforms corresponding to resonant ($h\nu=\SI{85}{\milli\electronvolt}$) and off-resonant ($h\nu=\SI{165}{\milli\electronvolt}$) phonon excitation for different fluences. The results are shown in Fig.~\ref{Fig:FluenceFreq}A. The frequency changes linearly with fluence in both magnetic phases for excitation in resonance with the phonon. For the off-resonant excitation, the frequency is nearly independent of the fluence. The corresponding time-dynamics from which these frequencies were extracted are shown in Fig.~\ref{Fig:FluenceFreq}B-E.

\FloatBarrier

\newpage

\subsection{Time-evolution analysis of the soft-mode frequency}

\begin{figure*}[h]
	\centering
	\includegraphics[width =0.8\linewidth]{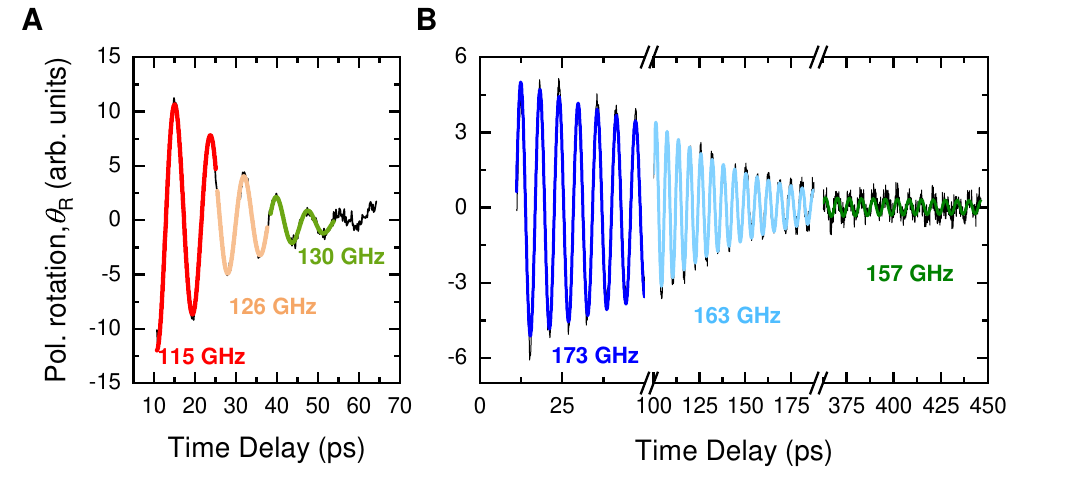}
	\caption{\textbf{Time-evolution of the spin precession frequencies.} Time-resolved trace of the spin precession dynamics in the (\textbf{A}) AFM phase  (\SI{46}{\kelvin}) and in the (\textbf{B}) WFM phase (\SI{55}{\kelvin}). The color lines are fits with exponentially damped oscillations applied to different time windows of the measured precession. The numbers below show the central frequency of these oscillations extracted from the fits.} 
	\label{FigureS5}
\end{figure*}

In order to reveal the dynamics of the altered spin precession frequency, we estimated its emergence and relaxation time, by performing a time-resolved analysis of the frequency of the spin oscillations. Figure~\ref{FigureS5}a,b show time-resolved traces of the spin oscillations in the AFM and WFM phases. The fits with damped sine function were applied to time slices of the magnetic dynamics at consecutive time-delays following the resonant pump excitation and reveal a gradual change in the frequency. To obtain a deeper insight into the dynamics of the frequency we reduced the length of the section to approximately half of the spin precession period. The ultimate resolution of this method is limited to a half-period of the spin-precession time which amounts to around \SI{5}{\pico\second}. The extracted values of the transient frequencies are shown in Fig.~2 in the main text. The uncertainty in x-values in this figure denote the width of the time window which was used for the fit. 

\newpage

\section*{S4 Phonon-driven control of the magneto-crystalline anisotropy}
\setcounter{subsection}{0}

\subsection{Selective control of parameters defining the anisotropy}

\begin{figure}
	\centering
	\includegraphics[width=\linewidth]{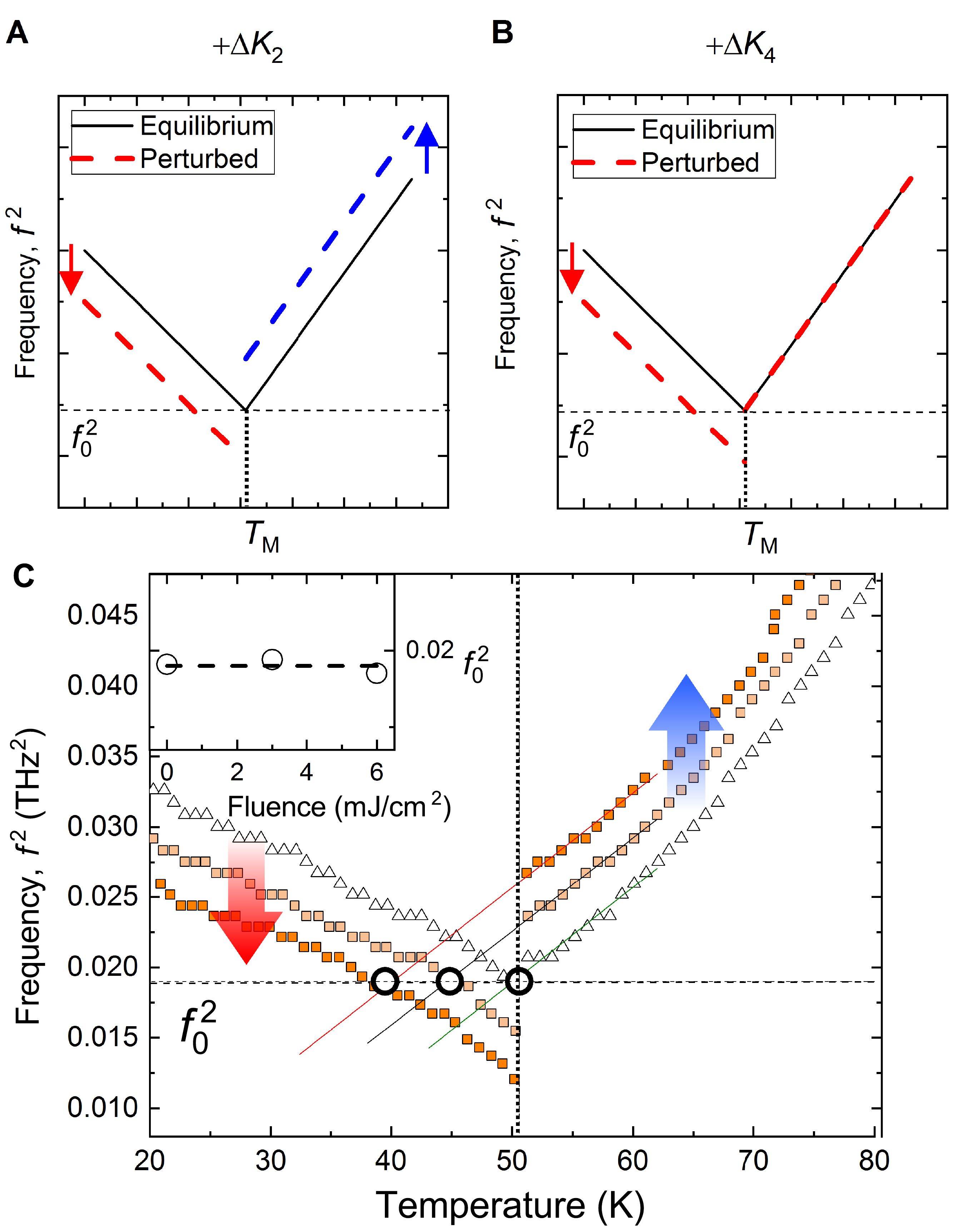}
	\caption{\textbf{Changing the soft-mode frequency.} (\textbf{A} and \textbf{B}) Schematic of the squared spin precession frequencies for the altered constants $K_2 + \Delta K_2$ (A) and $K_4 + \Delta K_4$ (B), indicated by dashed lines, following Eq.~\ref{Eq:ChangeConstants}. The temperature dependence of the spin precession frequency for the equilibrium values of the phenomenological constants $K_2$ and $K_4$ is indicated by the solid black line. (\textbf{C}) The extracted frequencies after excitation off-resonantly (black open triangles) and on-resonantly at different fluences (filled square markers). Intersection of the two frequency branches leads to the value $f_0^2$, which is shown in the inset for different pump fluences.
	} 
	\label{Fig:SupplTheory}
\end{figure}

In order to study the impact of resonant pumping of the lattice vibrational mode on the temperature dependence of the spin precession frequency, we initially assume that the resonant excitation changes one of the two anisotropy parameters. These anisotropy parameter describe the magnetic potential (Eq.~\ref{Eq:potential}.) We introduce renormalized (transient) anisotropy parameters for which the equilibrium values get an addition $\Delta K_2$ and $\Delta K_4$ such that: 
\begin{align}
\begin{split}
K_2\longrightarrow\;K_2+\Delta K_2 \\
K_4\longrightarrow\;K_4+\Delta K_4
\end{split}
\label{Eq:ChangeConstants}
\end{align}
Figure~\ref{Fig:SupplTheory}a~and~\ref{Fig:SupplTheory}b demonstrate the effect of each of these additives on the temperature dependence of the squared spin precession frequency $f^2$. A qualitative comparison of these plots with the experimental data shown~in Fig.~\ref{Fig:SupplTheory}C points that the phonon pumping mainly influences thee $K_2$ anisotropy parameter.

\subsection{Reconstructing the transient potential using spin precession}
\label{Sec:S5B}

As the curvature at the potential minimum defines the frequency of the soft mode oscillations, we used the frequency of the spin precession excited by the ISRS-mechanism (off-resonant excitation at $h\nu=\SI{165}{\milli\electronvolt}$) and by the phonon-pumping (resonant excitation at $h\nu=\SI{85}{\milli\electronvolt}$) to reconstruct the exact profile of the potential function before ($t<0$) and after ($t>0$) the phonon-pumping. First we determined the strength of the temperature independent $K_4$ parameter by exploiting the fact that at the transition temperature $T_\textnormal{M}$ the anisotropy parameters are coupled by the relation:
\begin{equation}
2K_2\left(T_\textnormal{M}\right) = K_4
\end{equation}
Hence, the temperature-independent value of the $K_4$ anisotropy can be extracted from the square of the angular frequency $\omega_0=2\pi f_0$ exactly at $T_\textnormal{M}$, using Eq.~\ref{Eq:Frequency} such that:
\begin{equation}
K_4 \left(T_\textnormal{M}\right)=2 \left(\frac{M_0}{\gamma}\right) \frac{\omega^2_0(T_\textnormal{M})}{\;\omega_\textnormal{exc}}
\label{Eq:K4}
\end{equation}
The value of $f_0^2$ for the oscillations excited by the lattice pumping can be found by intersecting the squared frequency branches corresponding to the AFM and WFM phases as exemplified in Fig.~\ref{Fig:SupplTheory}C. The inset in this figure shows that $f_0^2$ is independent of the laser fluence and equals $f_0^2$= \SI{0.0192}{\tera\hertz\squared}, matching the value of the $f_0^2$ in equilibrium.
This complies with our conclusions that the phonon-pumping does not affect $K_4$. We used the following material parameters to extract the value of the $K_4$ energy:

\begin{table}[htpb!]
	\caption{}
	\label{tab:Material_param}
	\begin{tabular}{ccccccc|c}
		\hline
		\hline
		& \multicolumn{5}{c|}{Parameters} & \multicolumn{1}{c}{Calculated value} \\ 
		\hline
		 $K_4$ (\SI{}{\erg\per\gram}) \cite{Balbashov1985} & $\omega_\textnormal{exc}$ (\si{THz}) 	&	$\rho$ (\SI{}{\g\per\centi\meter\cubed}) \cite{Balbashov1985} & M$_0$ ($\mu$eV/(T u.c.)) & $\gamma$ (THz/T) & $V_\textnormal{u.c.}$ (\SI{}{\angstrom 
			\cubed}) \cite{Treves1965} &  $K_4$ (\SI{}{\micro\electronvolt}/u.c.) \\
		\hline
		3.5$\cdot 10^4$ & 1565 & 8.0 & $1.16\cdot10^3$ & 0.028 & 262 &  40  \\
		
		\hline
	\end{tabular}
	%}
	\\
\end{table}
The value of $\omega_\textnormal{exc}$ was calculated to match the experimentally defined value of $K_4$ from Ref. \cite{Balbashov1985} with Eq.~\ref{Eq:K4}.

%Using the density $\rho$=\SI{8.0}{\g\per\centi\meter\cubed} and unit cell volume $V_\textnormal{u.c.}$=\SI{232}{\angstrom 
%	\cubed}, this gives $K_4$=\SI{40}{\micro\electronvolt}/u.c. Moreover, 
Once the value of the constant $K_4$ is known, the temperature dependence of the constant $K_2$ can be extracted using:

\begin{align*}
K_2\left(T\right)=
\frac{M_0}{\gamma}
\begin{cases}
\frac{1}{\;\omega_{\textnormal{exc}}}\omega^2(T)
,  &\text{  $T<T_\textnormal{M}$} \\      
K_4-\frac{1}{\;\omega_{\textnormal{exc}}}\omega^2(T), &\text{  $T>T_\textnormal{M}$} \\
\end{cases}
\end{align*}

\begin{figure}[H]
	\centering
	\includegraphics[width =\linewidth]{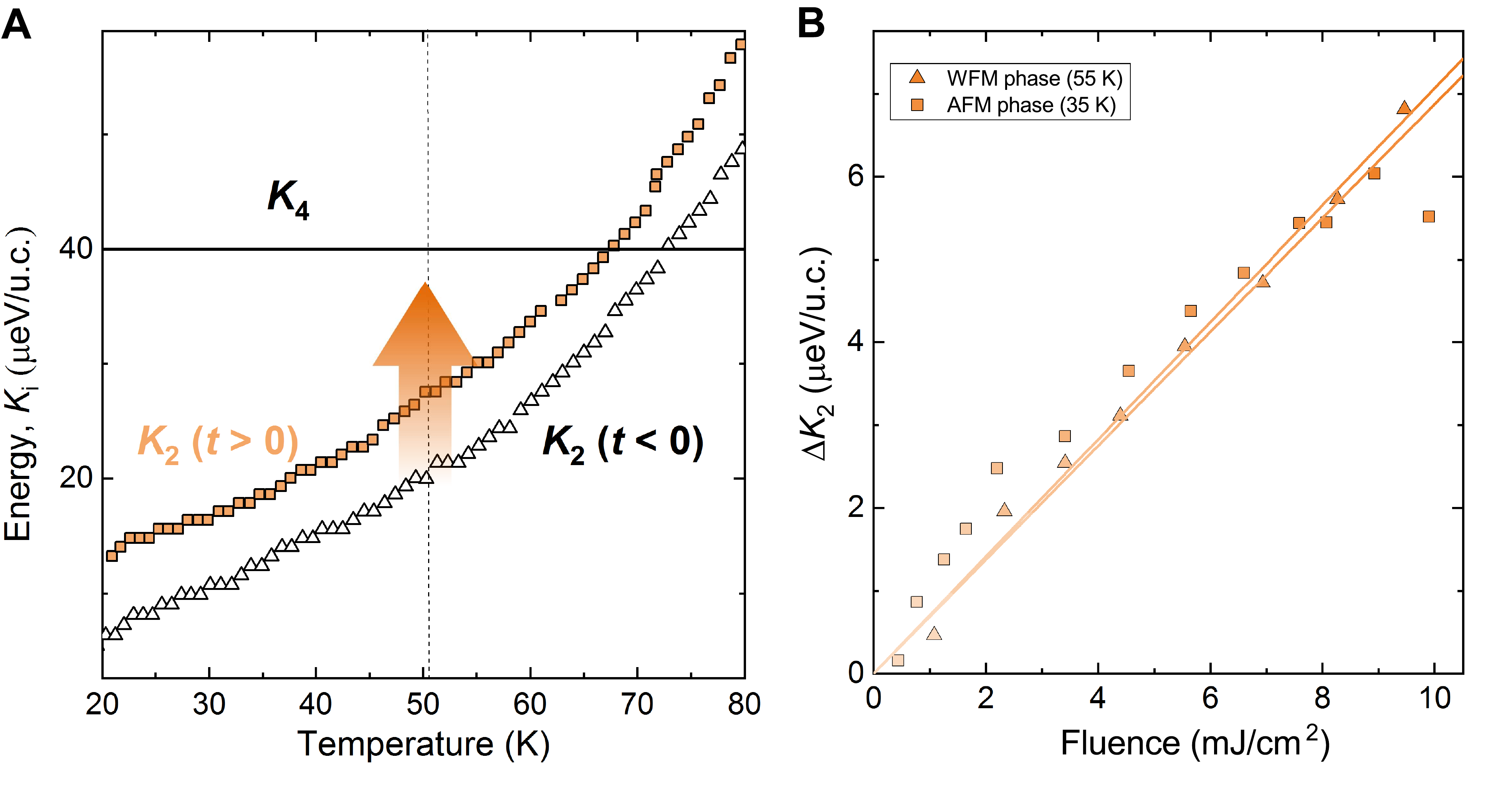}
	\caption{\textbf{Temperature and fluence dependence of the $K_2$ anisotropy.}
		(\textbf{A}) The anisotropy parameters $K_2$ and $K_4$ before and after the phonon pumping as extracted from the temperature dependence of the soft mode precession frequencies. (\textbf{B}) The phonon-induced additive $\Delta K_2$ to the $K_2$ anisotropy as a function of the pump fluence in the AFM (square markers) and the WFM (triangular markers) phase, with linear fits.
	} 
	\label{Fig:DeltaK2}
\end{figure}

\begin{figure}[H]
	\centering
	\includegraphics[width =\linewidth]{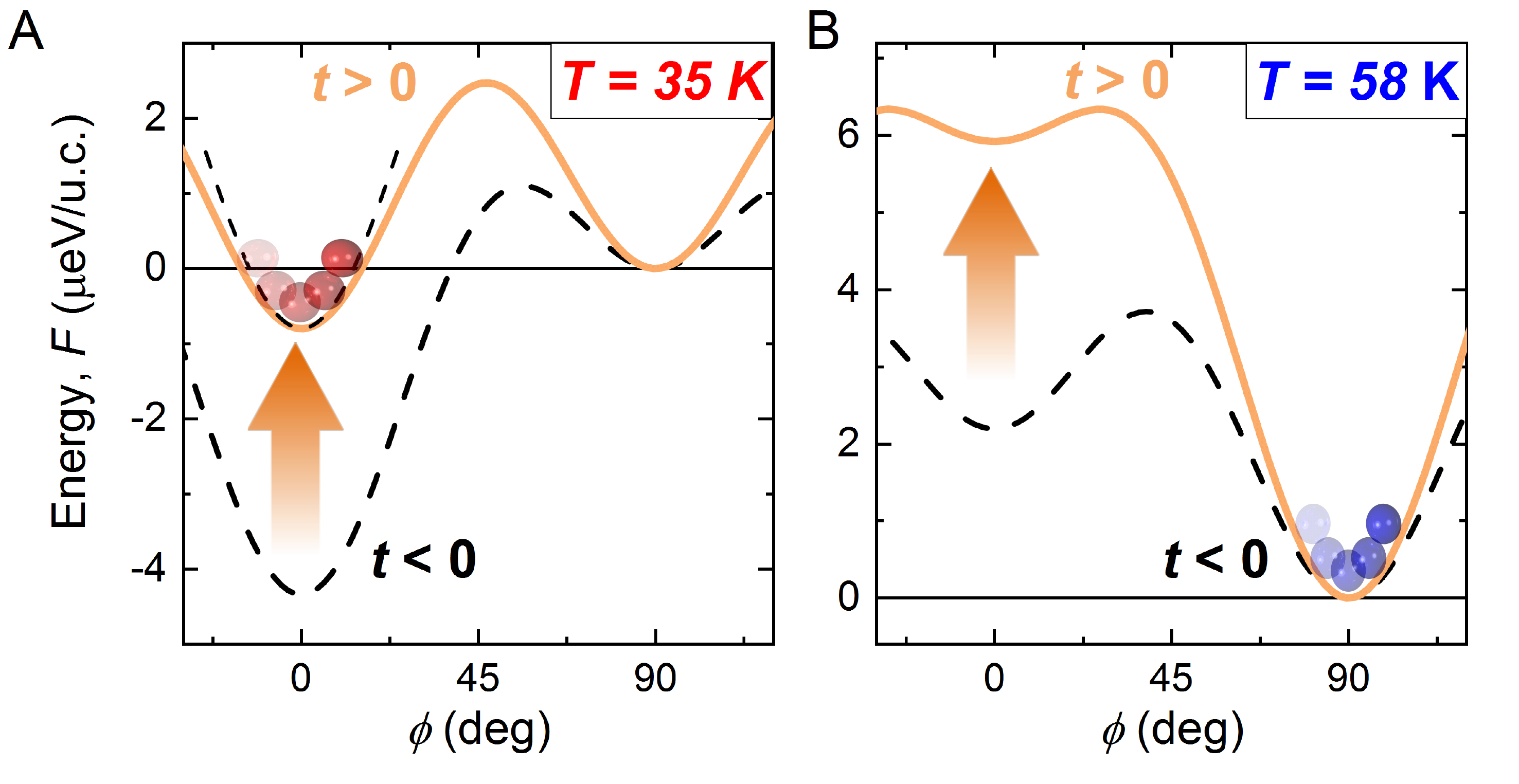}
	\caption{\textbf{Evolution of the magnetic potential after phonon-pumping.}	Equilibrium (black, dashed line, $t<0$) and reconstructed transient magnetic potential (orange line, $t>0$) after excitation of the phonon mode in the AFM (\textbf{A}) and WFM (\textbf{B}) phase. The position of the ball indicates the magnetic ground state of the system around which spin oscillations take place.
	} 
	\label{Fig:Metastability}
\end{figure}

The results are presented in Fig.~\ref{Fig:DeltaK2}A and confirm that the phonon pumping changes the value of $K_\textrm{2}$. The change $\Delta K_2$ is continuous at $T_\textnormal{M}$ and follows a linear dependence on the pump fluence,~see Fig.~\ref{Fig:DeltaK2}B. The evolution of the reconstructed magnetic potential before and after the excitation is shown in Fig.~\ref{Fig:Metastability}A and~\ref{Fig:Metastability}B, respectively. %The reconstruction clearly indicates that in proximity to the $T_\textrm{M}$ the phonon-pumping sets the system in the metastable state, see \ref{Fig:Metastability}d. The energy of this AFM state is higher than that of the WFM and for this reason this state cannot be realized in equilibrium, see~\ref{Fig:Metastability}c, but once induced in equilibrium it can be observed experimentally. This is due to the fact that changes in the magnetic potential are very fast faster than the nucleation. Thus the phonon-pumping presents a viable way to access such hidden state.   

%and can be as large as \SI{35}{\percent} at $T_\textnormal{M}$. %The Morin point defines the temperature at which the energies of the AFM and WFM phases are equal ($\Delta F=0$) and can be extracted from the relation $K_2(T_\textnormal{M})=\frac{1}{2}K_4$~\cite{Balbashov1985}. \textcolor{red}{The phonon-pumping increases the value of the $K_2$ parameter such that the intersection with $\frac{1}{2}K_4$ occurs at a lower temperature $T_\textnormal{M}^{'}$. This value can be considered as an effective dynamical Morin temperature leading to a dynamical shift of the Morin temperature $\Delta T_\textnormal{M}$ to lower temperatures, see inset in Fig.~\ref{Fig:DeltaK2}.}

\subsection{Effect of the laser fluence on the spin precession frequency and the anisotropy constant $K_2$}
\label{Sec:S4C}

%Figure~\ref{Fig:DeltaK2}b shows that $\Delta K_2$, the pump-induced additive to the anisotropy parameter $K_2$, linearly depends on fluence of the laser pump. 
The linear dependence of $\Delta K_2$ on the pump fluence signals a quadratic dependence on the amplitude of the electric field of the pump pulse. This can be easily understood if we consider the expansion of $\Delta K_2$ in a series of the normal phonon coordinate $Q_{\textnormal{IR}}$:

%the frequency of the spin precession depends on the fluence of the pump pulse. It is seen that for the off-resonant excitation the central frequency of the spin precession in both magnetic phases remains constant upon the fluence increase, indicating a perturbative regime of the excitation. As discussed in the main text, pumping at the phonon resonance induces large changes in the frequency of the spin precession. The curves follow an approximately linear dependence, with opposite slopes of nearly equal magnitude for two magnetic phases.
%The linear dependence of the frequency shifts results in linear dependence of the anisotropy constant $\Delta K_2$ as a function of the laser fluence, see Fig.~\ref{Fig:FigureS4}b. 

\begin{equation}
\Delta K_2=K_2(Q_{\textnormal{IR}})-K_2=A_1 Q_{\textnormal{IR}}+A_2 Q_{\textnormal{IR}}^2+\ldots, 
\label{Eq:anisot}
\end{equation}
with $A_{1,2}$ being phenomenological constants. The phonon coordinate is defined as $Q_{\textnormal{IR}}=Q_0\sin\left(2\pi\Omega_{\textnormal{IR}}t\right)$, where $Q_0$ and $\Omega_{\textnormal{IR}}$ are the amplitude and the frequency of the infrared-active phonon, respectively. 
Since the frequency of the infrared-active phonon is significantly higher than the one of the spin precession ($\Omega_{\textnormal{IR}}\gg f$), Eq.~\ref{Eq:anisot} has to be time-averaged. % to account for the effect of the excitation of the phonon vibration on the magnetic precession.
The averaging shows that the phonon-driven additive $\Delta K_2$ is determined by the rectified part of the phonon-coordinate which depends quadratically on the amplitude $Q_0$, such that:

\begin{equation}
\Delta K_2=\frac{1}{2}B Q_0^2. 
\end{equation}
Thereby it explains the observed quadratic dependence of the $\Delta K_2$ additive on the amplitude of the driving electric field. To explain a possible physical origin of this quadratic dependence we performed a set of the DFT simulations to find micorscopic sources of such nonlinear lattice-driven response.
\newpage	

\section*{S5. Nonlinear lattice dynamics and their effect on magnetic interactions}
\label{Sec:S8DFT}
\setcounter{subsection}{0}
\subsection{Theory DFT simulations}

\subsubsection{Technical details}

We simulated the $Pnma$ phase of \ce{DyFeO3} through density functional theory (DFT) \cite{Hohenberg1964,Kohn1965} and its projected augmented wave (PAW)
\cite{Blochl1994} flavour as implemented in the ABINIT package \cite{Gonze2002, abinit2016,Torrent2008}. The PAW atomic potentials used for \ce{Fe} and \ce{O} were downloaded from the GBRV pseudopotential library \cite{Garrity2014} with $3s$, $3p$, $3d$, $4s$ and $2s$, $2p$ considered as valence states for \ce{Fe} and \ce{O} respectively.
For \ce{Dy} we used the PAW atomic potential from M. Topsakal et al. \cite{Topsakal2014}, where  the valence states are $4f$, $5s$, $5p$ and $6s$ (with $f$ electrons in the valence).
We used the PBEsol GGA functional \cite{Perdew2008} for the exchange correlation interaction and Hubbard corrections \cite{Liechtenstein1995} on both \ce{Fe} and \ce{Dy} of respectively \SIrange[range-units = single, range-phrase = {~and~}]{4}{7}{\electronvolt} have been used to have the closest possible properties with respect to experiments (lattice constants). 
All the calculations were done with a 6$\times$6$\times$4 mesh of k-points for the reciprocal space and a cut-off energy on the plane wave expansion of 36 Hartree and 72 Hartree for the second finer grid inside the PAW spheres.
To calculate the phonons, the frozen phonon technique has been used through the phonopy software \cite{Togo2015}.

%In order to determine the ground state for the Dy-$f$ orbital occupations, we constrained different occupation matrices at the beginning of each calculation following the scheme reported in Ref. \cite{dorado}. All the results where the f-electrons are treated as valence states are those with this ground state occupation of the $f$ orbitals%, which we kept for all the different calculations (relaxation, frozen phonon and condensed phonons).
%We found that the lowest energy $f$ orbital occupations case is also the one that gives the best agreement with the experimental cell parameters, which confirms our choice of the $U$ parameter (see table \ref{tab:lat_param}).
%The calculations of the phonon frequencies of which the results are presented below, were performed treating the $f$ electrons as core electrons. The calculations of the phonon-phonon coupling and the magnetic interactions were performed with the $f$ electrons treated as valence states.

\begin{table}[htbp!]
	\caption{Calculated \ce{DyFeO3} phonon IR active mode frequency (\si{\milli\electronvolt}).
		The first column gives the mode label  of the $Pnma$ structure with $f$ electrons in the core and the second column the calculated frequency. %The third column presents the mode label in the $Pna2_1$ space group and the fourth column the corresponding calculated frequencies with $f$ electrons in the valence. The last column ($\Delta$) gives the difference in frequency between $f_{core}$ and $f_{valence}$.}
	}
	\label{IR active modes}
	%\resizebox{7cm}{!}{
	\begin{tabular}[t]{cHcHcH@{\hspace*{-\tabcolsep}}}
		\hline
		\hline
		Phonon mode	&	$f_\textnormal{IR}$ (\si{\milli\electronvolt})	& &	label($Pna2_1$)	&	$\Omega_\textnormal{IR}$ (\si{\milli\electronvolt})	&	 $\Delta$	\\
		\hline
		\hline
		
		B$_{1u}$(1)	&	13.51	&\  \ \ &	B$_2$(6)	&	13.14	&	0.37		\\
		B$_{1u}$(2)	&	20.83	&&	B$_2$(7)	&	21.70	&	-0.87		\\
		B$_{1u}$(3)	&	30.25	&&	B$_2$(8)	&	31.00	&	-0.74		\\
		B$_{1u}$(4)	&	33.97	&&	B$_2$(9)	&	34.47	&	-0.50		\\
		B$_{1u}$(5)	&	40.17	&&	B$_2$(10)	&	39.92	&	0.25		\\
		B$_{1u}$(6)	&	42.90	&&	B$_2$(11)	&	43.02	&	-0.12		\\
		B$_{1u}$(7)	&	49.47	&&	B$_2$(12)	&	48.72	&	0.74		\\
		B$_{1u}$(8)	&	62.11	&&	B$_2$(13)	&	61.62	&	0.50		\\
		B$_{1u}$(9)	&	67.07	&&	B$_2$(14)	&	67.32	&	-0.25		\\
		\hline
		B$_{2u}$(1)	&	18.35	&&	A$_1$(9)	&	18.84	&	-0.50		\\
		B$_{2u}$(2)	&	19.71	&&	A$_1$(10)	&	20.08	&	-0.37		\\
		B$_{2u}$(3)	&	31.37	&&	A$_1$(11)	&	32.23	&	-0.87		\\
		B$_{2u}$(4)	&	37.07	&&	A$_1$(12)	&	37.19	&	-0.12		\\
		B$_{2u}$(5)	&	43.15	&&	A$_1$(13)	&	43.52	&	-0.37		\\
		B$_{2u}$(6)	&	59.51	&&	A$_1$(14)	&	60.75	&	-1.24		\\
		B$_{2u}$(7)	&	63.85	&&	A$_1$(15)	&	65.34	&	-1.49		\\
		\hline
		B$_{3u}$(1)	&	12.03	&&	B$_1$(6)	&	12.40	&	-0.37		\\
		B$_{3u}$(2)	&	23.43	&&	B$_1$(7)	&	23.18	&	0.25		\\
		B$_{3u}$(3)	&	28.27	&&	B$_1$(8)	&	29.01	&	-0.74		\\
		B$_{3u}$(4)	&	35.83	&&	B$_1$(9)	&	35.95	&	-0.12		\\
		B$_{3u}$(5)	&	38.06	&&	B$_1$(10)	&	38.31	&	-0.25		\\
		B$_{3u}$(6)	&	41.29	&&	B$_1$(11)	&	41.16	&	0.12		\\
		B$_{3u}$(7)	&	52.69	&&	B$_1$(12)	&	52.57	&	0.12		\\
		B$_{3u}$(8)	&	63.23	&&	B$_1$(13)	&	63.73	&	-0.50		\\
		B$_{3u}$(9)	&	64.84	&&	B$_1$(14)	&	65.09	&	-0.25		\\
		\hline
	\end{tabular}
	\\
\end{table}

\begin{table}
	%\captionsetup{font=scriptsize}
	\caption{ Calculated \ce{DyFeO3} Raman active mode frequency (\si{\milli\electronvolt}).  In the first column we show the mode label within the $Pnma$ structure and in the second column we report the frequency of the mode. %with $f$ electrons in the core. 
	In the third and fourth column we also report, respectively, the calculated frequencies and the experimental measurements reported by Weber et al. in Ref.~\protect{\cite{Weber2016}}.}
	\label{Raman active modes}
	%\resizebox{7cm}{!}{
	\setlength{\tabcolsep}{12pt}
	\begin{tabular}{cHHcccH@{\hspace*{-\tabcolsep}}}
		\hline
		\hline
		Phonon mode	&	$f_\textrm{R}$	&	label($Pna2_1$)	&	$\Omega_\textnormal{R} (\textrm{meV})$	&	Ref.	&	Exp.	&	$\Delta$	\\
		\hline
		\hline
		
		A$_{g}$(1)	&	13.51	&	A$_1$(1)	&	13.27	&	13.89	&	14.01	&	0.25		\\
		A$_{g}$(2)	&	15.99	&	A$_1$(2)	&	16.49	&	16.74	&	17.36	&	-0.50		\\
		A$_{g}$(3)	&	31.61	&	A$_1$(3)	&	30.87	&	32.48	&	32.36	&	0.74		\\
		A$_{g}$(4)	&	40.91	&	A$_1$(4)	&	40.67	&	41.16	&	42.28	&	0.25		\\
		A$_{g}$(5)	&	49.72	&	A$_1$(5)	&	50.58	&	52.32	&	52.32	&	-0.87		\\
		A$_{g}$(6)	&	51.45	&	A$_1$(6)	&	50.58	&	51.45	&	51.70	&	0.87		\\
		A$_{g}$(7)	&	49.72	&	A$_1$(7)	&	51.20	&	52.32	&	52.32	&	-1.49		\\
		A$_{g}$(8)	&	59.39	&	A$_1$(8)	&	59.51	&	60.75	&	61.49	&	-0.12		\\		  \hline
		B$_{1g}$(1)	&	15.99	&	B$_1$(1)	&	16.86	&	16.74	&	--	&	-0.87		\\
		B$_{1g}$(2)	&	29.88	&	B$_1$(2)	&	30.38	&	31.00	&	--	&	-0.50		\\
		B$_{1g}$(3)	&	43.52	&	B$_1$(3)	&	43.76	&	44.51	&	44.76	&	-0.25		\\
		B$_{1g}$(4)	&	51.08	&	B$_1$(4)	&	51.95	&	52.94	&	52.94	&	-0.87		\\
		B$_{1g}$(5)	&	72.53	&	B$_1$(5)	&	73.27	&	73.52	&	--	&	-0.74		\\
		\hline
		B$_{2g}$(1)	&	13.02	&	A$_2$(1)	&	13.27	&	13.51	&	13.76	&	-0.25		\\
		B$_{2g}$(2)	&	19.46	&	A$_2$(2)	&	19.22	&	19.96	&	20.21	&	0.25		\\
		B$_{2g}$(3)	&	38.06	&	A$_2$(3)	&	36.95	&	38.56	&	40.29	&	1.12		\\
		B$_{2g}$(4)	&	42.90	&	A$_2$(4)	&	43.15	&	43.52	&	--	&	-0.25		\\
		B$_{2g}$(5)	&	59.39	&	A$_2$(5)	&	58.77	&	59.76	&	61.25	&	0.62		\\
		B$_{2g}$(6)	&	65.34	&	A$_2$(6)	&	65.34	&	66.21	&	--	&	0.00		\\
		B$_{2g}$(7)	&	75.26	&	A$_2$(7)	&	75.63	&	75.88	&	77.36	&	-0.37		\\
		\hline
		B$_{3g}$(1)	&	14.51	&	B$_2$(1)	&	14.88	&	15.25	&	--	&	-0.37		\\
		B$_{3g}$(2)	&	37.69	&	B$_2$(2)	&	38.06	&	38.56	&	--	&	-0.37		\\
		B$_{3g}$(3)	&	50.58	&	B$_2$(3)	&	51.70	&	52.57	&	53.68	&	-1.12		\\
		B$_{3g}$(4)	&	56.16	&	B$_2$(4)	&	56.41	&	57.03	&	58.77	&	-0.25		\\
		B$_{3g}$(5)	&	77.98	&	B$_2$(5)	&	78.98	&	78.98	&	79.22	&	-0.99		\\
		\hline
	\end{tabular}
	%}
	\\
\end{table}

%In order to determine the ground state for the Dy-$f$ orbital occupations, we constrained different occupation matrices at the beginning of each calculation following the scheme reported in Ref. \cite{Dorado2009}. 
%We found that the lowest energy $f$ orbital occupations case is also the one that gives the best agreement with the experimental cell parameters, which confirms our choice of the $U$ parameter (see table \ref{tab:lat_param}).
%All the results where the f-electrons are treated as valence states are those with this ground state occupation of the $f$ orbitals, which we kept for all the different calculations (relaxation, frozen phonon and condensed phonons).

%\begin{table}[htpb!]
%\caption{lattice parameters and band gaps.}
%\label{tab:lat_param}
%\begin{tabular}{cccccc}
%\hline
%\hline
%&  a	&	b	& c & Band-gap(eV) \\
%\hline
%Exp \cite{Weber2016} & 5.302 & 5.598 & 7.623 & -- \\
%DFT($f_{valence}$) & 5.292 & 5.594 & 7.615 & 2.30 \\
%DFT($f_{core}$) & 5.247 & 5.588 & 7.588 & 2.28 \\
%
%\hline
%\end{tabular}
%%}
%\\
%\end{table}

\subsubsection{Phonons}
%To calculate the phonons, we used both PAW's where the $f$ electron considered as valence and as core. The results were very close so we kept the $f$ electrons in core for phonon calculations.\\
In Tables \ref{IR active modes} and \ref{Raman active modes} we report calculated IR and Raman-active mode frequencies. %The main difference between these two calculations is that with $f_{valence}$ \ce{Dy} is magnetic while with $f_{core}$ \ce{Dy} atoms do not possess a magnetic moment.
Both the Raman and IR active modes are in good agreement with the experiments and previous DFT calculations done by Weber et al. \cite{Weber2016}. 
The few small deviations between our results and the previous calculations  can be due to the use of different lattice parameters (experimental cell parameters were used by Weber et al. while we used relaxed PBEsol cell parameters), to the different PAW atomic potentials or because we used f-electrons in the valence for Dy. 

\subsubsection{Phonon-phonon coupling model}

%In order to understand to which phonon modes the laser pulse couples, we considered the laser field to be a Gaussian function with a distribution given by Eq.\ref{eq:laser}. 
%In this equation $\sigma_{\omega}$ gives the frequency spread of the electric field and E$_0$ gives the peak of the pulse while $\omega_0$ gives its mean frequency.
%
%
%\begin{equation} \label{eq:laser}
%E(\omega) = \frac{E_0 }{2\pi\sigma_{\omega}}e^{\frac{-(\omega-\omega_{0})^2}{2\sigma_{\omega}}}
%\end{equation}
%
%In Fig.\ref{fig:laser_pls} we report the laser pulse distribution and

\begin{figure}[h]
	\centering
	\includegraphics[width=\linewidth, ,keepaspectratio=true]{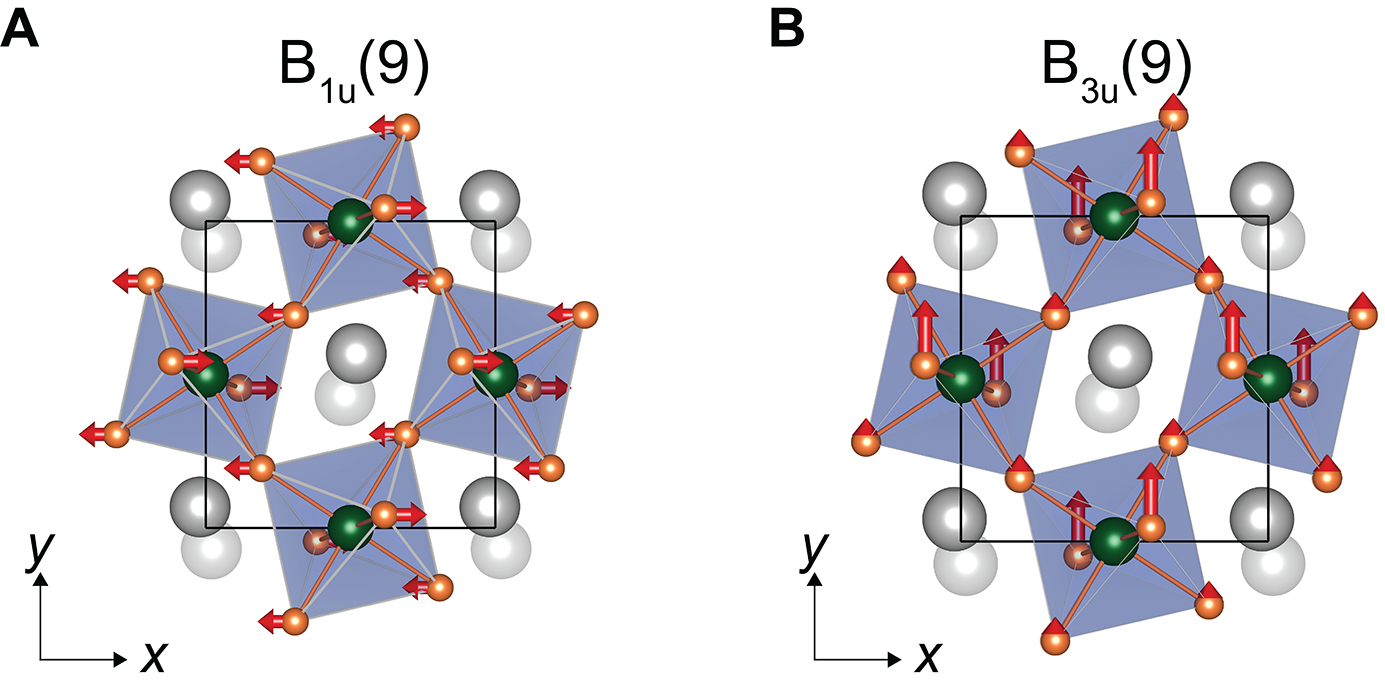} 
	\caption{\textbf{Highest frequency phonon modes} Schematic picture of the polar component of the eigenvectors of the highest-frequency infrared-active, polarized orthogonally $B_\textnormal{1u}$(9) (A) and $B_\textnormal{3u}$(9) (B) phonon modes. Grey/green/orange spheres represent dysprosium/iron/oxygen atoms.}
	\label{Fig:IR_modes}
\end{figure}

In our experiment the laser pulse excites the B$_{1u}$(9) (Fig.~\ref{Fig:IR_modes}A) and B$_{3u}$(9) (Fig.~\ref{Fig:IR_modes}B) modes having the highest frequencies. These two modes are polar in the $x$ and $y$-directions and correspond to a stretching of the oxygen bonds.

%\begin{figure}
%\centering
%\includegraphics[width=0.5\linewidth, ,keepaspectratio=true]{Figures_supp/laser-pulse.png} 
%\caption{Spectrum of the laser pulse with $\omega_0$ at \SI{85}{\milli\electronvolt} and a full width at half maximum ($ \sigma_{\omega}$) of \SI{9}{\milli\electronvolt}, as used in the experiment. We also show the B$_{1u}$ and B$_{3u}$ mode frequency positions as dashed red and blue vertical lines respectively. 
%We can see that the laser pulse can excite mainly the highest frequency $B_{1u}$ mode at \SI{67}{\milli\electronvolt}.}
%\label{fig:laser_pls}
%\end{figure}
We studied different possibilities of phonon-phonon coupling that can cause a change in the magnetic interactions. 

One possibility is the nonlinear coupling of IR modes to other phononic modes, dubbed nonlinear phononics~\cite{Mankowsky2014,Forst2011}. According to this mechanism when large-amplitude atomic vibrations are excited, a non-linear response sets in, causing a net shift of the lattice energy potential along a (Raman- or Infrared-active) normal mode coordinate for a time given by the lifetime of the infrared-active phonon. The ensuing net distortion can consequently change the magnetic interactions of \ce{Dy} and \ce{Fe} and result in spin reorientation. To determine the relevant nonlinear phonon-phonon couplings, we used the potential energy $V(Q)$ containing several anharmonic phonon-phonon interactions:

\begin{equation} \label{eq:potential1}
\begin{split}
V(Q)=\omega_\textnormal{IR}^{2}Q_\textnormal{IR}^{2}+\omega_\textnormal{R}^{2}Q_\textnormal{R}^{2}+ C_\textnormal{R}Q_\textnormal{R}^3+\gamma_1Q_\textnormal{R}Q_\textnormal{IR}^2+\gamma_2Q_\textnormal{R}^2Q_\textnormal{IR}^2+\frac{1}{4}d_\textnormal{IR}Q_\textnormal{IR}^4+\frac{1}{4}d_\textnormal{R}Q_\textnormal{R}^4\\
\end{split}
\end{equation}

In Eq.~\ref{eq:potential1}, $Q_\textnormal{X}$ are mode amplitudes of phonon mode X (X = IR,R) with eigenfrequencies $\omega_\textnormal{X}$. $C_\textnormal{R}$, $\gamma_1$ , $d_\textnormal{IR}$ , $d_\textnormal{R}$ and $\gamma_2$  are coefficients fitted against DFT data in the spirit of Ref.~\cite{Juraschek2017}.

%To obtain the coupling constants in we fitted
% :   which we fitted against DFT calculations 
%Using PAW's for Dy with $f$ in core and in valence, we found that the differences are small so, we continued the fitting using the PAW's with $f$ in core.

We have considered the highest IR frequency B$_{1u}$(9) mode %(noted IR in Eq. 2) 
excited by the laser and its couplings to each A$_g$ mode
because they are the first order couplings coming into play in the 20 atoms perovskite $Pnma$ phase. The A$_g$ mode label being invariant under all the symmetry operations of the crystal, any coupling order with this mode is allowed such that the lowest order coming is of the type $Q_{\textnormal{A}_\textnormal{g}}Q_{\textnormal{B}_\textnormal{1u}}^2$.
As one can see in Table \ref{tab:Pot1}, of all the coupling orders, the $\gamma_1$ parameter gives the strongest coupling with the B$_{1u}$ mode.

\begin{table}[htbp!]
	\caption{DFT fitted coupling coefficients of Eq.\ref{eq:potential1} for B$_{1u}$(9) coupling with A$_g$(1) and A$_g$(2) modes. The units are ($\textnormal{meV}/(\sqrt{\textnormal{amu}}A^0)^\textnormal{n}$).}
	\label{tab:Pot1}
	\setlength{\tabcolsep}{7pt}
	\begin{tabular}{cccccc}
		\hline
		\hline
		Coupled modes &  C$_R$	&	d$_{IR}$	& d$_R$ & $\gamma_1$  & $\gamma_2$ \\
		\hline
		B$_{1u}$--A$_g$(1) & -0.004 & 0.0072 & 0.000 & 0.0681 & 0.000 \\
		B$_{1u}$--A$_g$(2) & 0.003 & 0.0072 & 0.000 & 0.1246 & 0.000 \\
		
		\hline
	\end{tabular}
\end{table}

\begin{figure}[h]
	\centering
	\includegraphics[width=\linewidth, ,keepaspectratio=true]{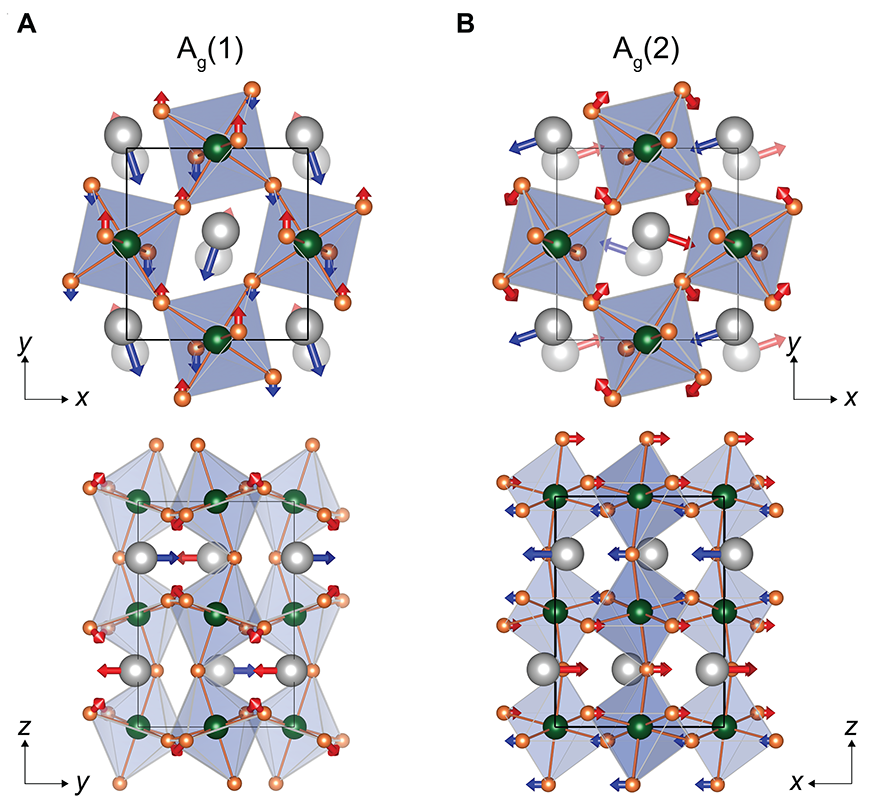} 
	\caption{\textbf{Lowest-frequency A$_g$ phonon modes.} Schematic picture of the eigenvectors of the $A_\textnormal{g}$(1) (A) and $A_\textnormal{g}$(2) (B) phonon modes, as viewed from two different directions. Grey/green/orange spheres represent dysprosium/iron/oxygen atoms.}
	\label{Fig:Ag_modes}
\end{figure}

After having fitted the coupling coefficients between the B$_{1u}$(9) mode and the A$_g$ modes, we found that the second lowest frequency A$_g$(2) mode (see its schematic representation in Fig. \ref{Fig:Ag_modes}B) gives the strongest coupling $\gamma_1$, closely followed by the first one A$_g$(1) (see Fig.~\ref{Fig:Ag_modes}A.) 

This also shows that considering the lowest A$_g$ mode frequency by default is not necessarily physically relevant since, for small frequency differences, if the coupling parameter $\gamma_1$ is larger for the second highest frequency mode than for the first one, it can dominate the dynamics of the system. 

The main difference between the A$_g$(1) and A$_g$(2) modes is in the way the antipolar motions of \ce{Dy} ions are involved.
Using the high symmetry reference cubic perovskite we can observe that the A$_g$(1) mode is dominated by the cubic X point antipolar motion mode of \ce{Dy} atoms (with coordinate [1/2, 0, 0] of the cubic Brillouin zone and symmetry label X$_5^-$), while the A$_g$(2) mode is dominated by the R point mode (coordinate [1/2, 1/2, 1/2] of the cubic Brillouin zone and symmetry label R$_4^-$). 
By doing the same analysis for the B$_{1u}$(9) mode, we found that this polar mode is dominated by the M$_2^+$ cubic mode label  involving oxygen octahedral  rotations ($a^0a^0b^+$ in Glazer notation \cite{Glazer1972}) plus the polar mode label $\Gamma_4^-$ (with polarization direction along the $x$ direction) plus R$_4^-$ antipolar motions of the \ce{Dy} atoms. Note that the Dy atoms have a very small amplitude contribution to the mode eigenvector due to the fact that at high frequency they are much less involved in the phonon vibrations.
The fact that the B$_\textnormal{1u}$(9) allows, by symmetry, antipolar motions of \ce{Dy} of the same type as the A$_g$(2) mode (R$_4^-$ label for both) can explain why this A$_g$(2) mode has a larger coupling with the B$_\textnormal{1u}$(9) mode than the A$_g$(1) mode. 

\subsubsection{Magnetic interactions}
\begin{figure*}[htb!]
	\centering
	\includegraphics[width=0.8\linewidth ,keepaspectratio=true]{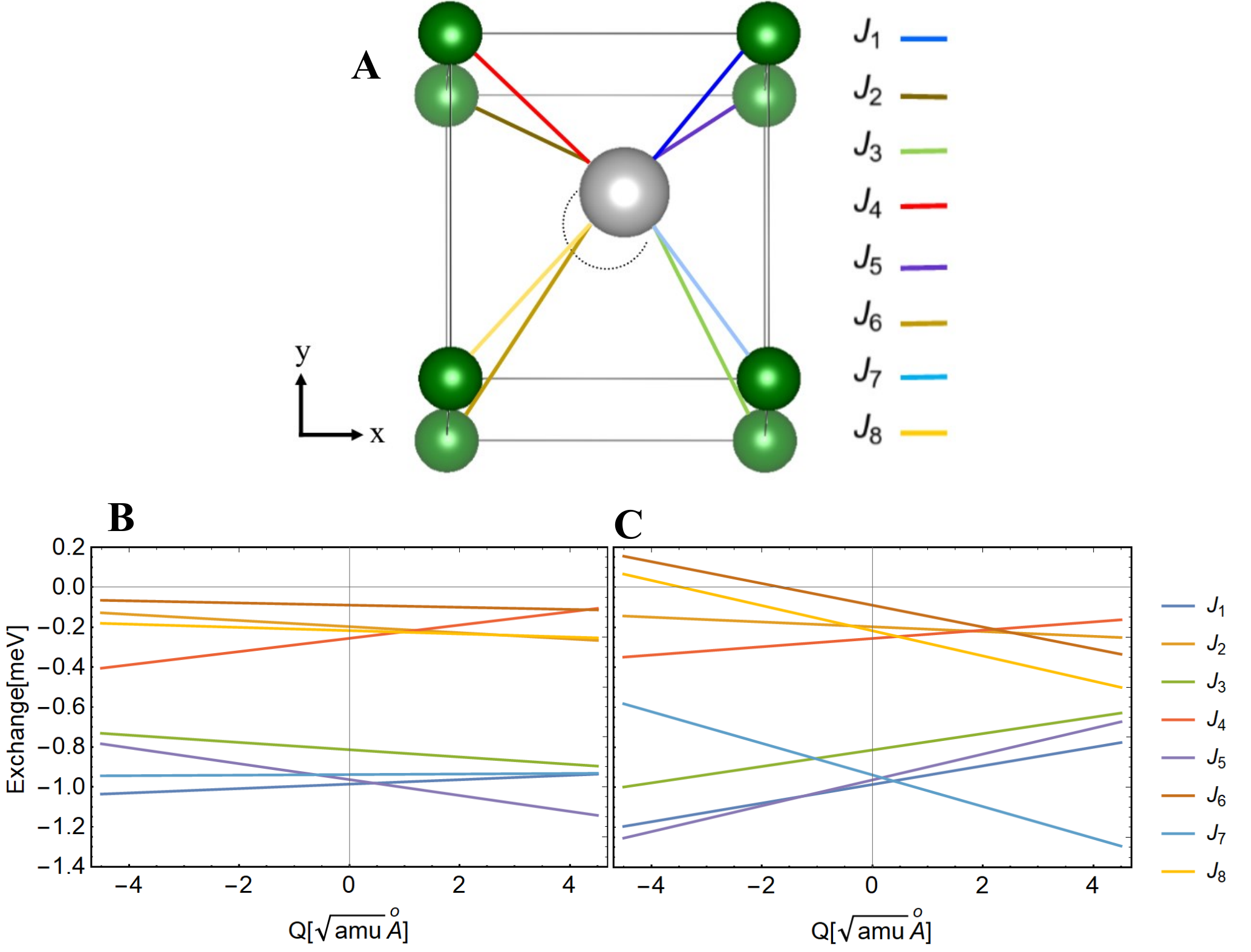} 
	\caption{\textbf{Dy-Fe superexchange as function of A$_g$ mode amplitudes.} (\textbf{A}) The Fe-Dy superexchange as decomposed into eight contributions related to the eight Fe ions surrounding one Dy ion. The gray ball corresponds to \ce{Dy} and the green balls correspond to \ce{Fe}; The empty dashed black circle corresponds to the equilibrium \ce{Dy} atom as if not displaced from the high-symmetry cubic position. (\textbf{B} and \textbf{C}) Superexchange interactions between \ce{Dy} and \ce{Fe} as a function of the distortions along the (B) A$_g$(1) mode and (C) A$_g$(2) mode coordinate (color scheme corresponds to the \ce{Dy}-\ce{Fe} bond color of panel A).}
	\label{fig:mag_int}
\end{figure*}

To study how the indirectly driven A$_g$ modes can change the magnetic interaction between \ce{Dy} and \ce{Fe}, we used a Green's function method to calculate the superexchange interaction as a function of the A$_g$ modes amplitude following the scheme of Korotin et al. \cite{Korotin2015}. We report the evolution of the \ce{Dy}-\ce{Fe} and \ce{Fe}-\ce{Fe} superexchanges with respect to both A$_g$(1) and A$_g$(2) mode amplitudes. 

We observe that the two Raman-active modes can drive different modifications of the Fe-Dy superexchange, which can be decomposed into eight contributions related to the eight Fe ions surrounding one Dy ion as shown in Fig.~\ref{fig:mag_int}A. In Fig.~\ref{fig:mag_int}B,C it is shown that for example, $J_3$ and $J_7$ have opposite trends with respect to both A$_g$(1) and A$_g$(2) mode condensation, while all the other $J_\textnormal{i}$ have the same trend. We also note that the A$_g$(2) mode has a larger effect on the $J$ than the A$_g$(1) mode (a larger slope of the linear $J$ vs mode amplitude trend). Table~\ref{tab:Dy-Fe} summarizes the linear variation of exchange with respect to the amplitude of the mode ($\mathrm{d}J/J\mathrm{d}Q$) for both \ce{Fe}-\ce{Dy} and \ce{Fe}-\ce{Fe} superexchanges. The A$_g$ modes affect much more the \ce{Fe}-\ce{Dy} exchange than the \ce{Fe}-\ce{Fe} one, the \ce{Fe}-\ce{Fe} exchange variation being negligible.
\begin{table}[htbp!]
	\caption{The normalized variation $\mathrm{d}J_i/(J_i\mathrm{d}Q)$ of the Fe-Dy and Fe-Fe exchange interactions as a function of the normal phonon coordinate written for A$_g$(1) and A$_g$(2) modes.}
	\label{tab:Dy-Fe}
	\begin{tabular}{ccc}
		\hline
		\hline
		Exchange & A$_g$(1),\;\%    & A$_g$(2),\;\% \\
		\hline
		J$_1$  &	-0.011  &	-0.047 \\
		J$_2$  &	0.077  &	0.060 \\
		J$_3$  &	0.022  &	  -0.050 \\
		J$_4$  &	-0.129  &	-0.081 \\
		J$_5$  &	0.041  &	-0.067 \\
		J$_6$  &	0.059  &	0.607 \\
		J$_7$  &	-0.001  &	0.084 \\
		J$_8$  &	0.037  &	0.289 \\
		\hline
		\hline
		Fe-Fe  &	0.001  &	0.001 \\
		\hline
	\end{tabular}
\end{table}

\begin{figure}[h]
	\centering
	\includegraphics[width=1\linewidth, ,keepaspectratio=true]{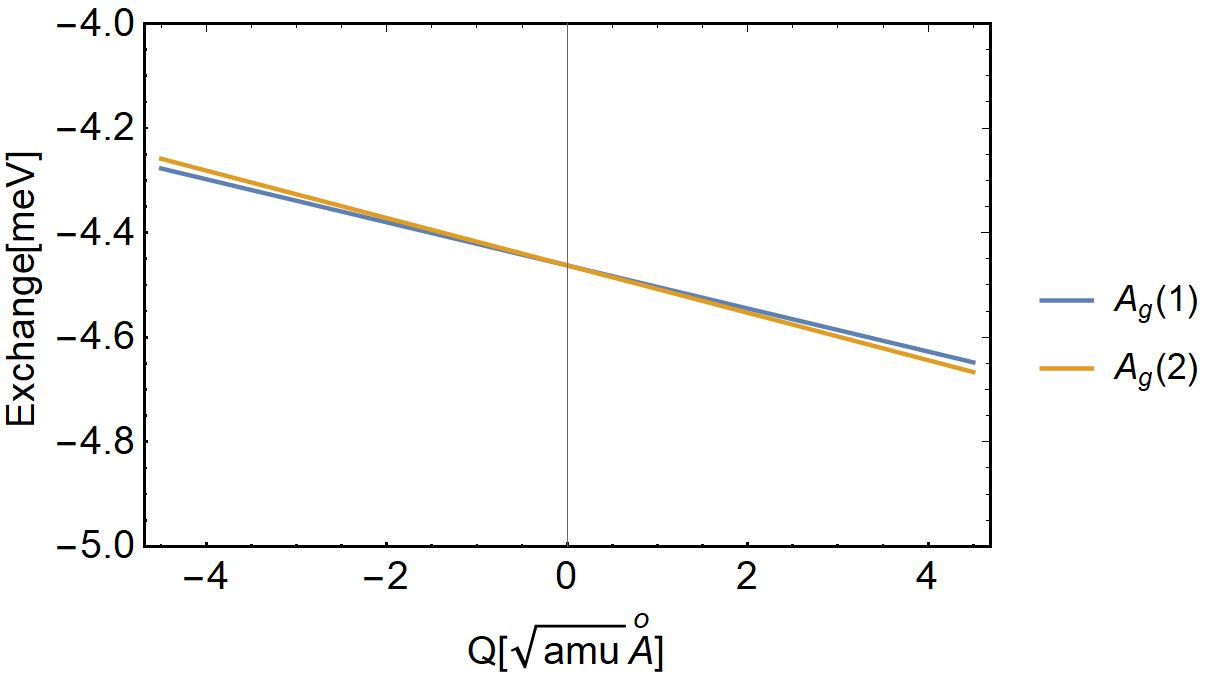} 
	\caption{\textbf{Dy-Fe superexchange vs. A$_g$ mode amplitude.} Change in the net superexchange interaction between each Dy ion and its eight Fe neighbours for A$_g$(1) (blue line) and A$_g$(2) (brown line). The result is obtained by summing up the eight contributions shown in Fig.~\ref{fig:mag_int}}
	\label{Fig:Tot_exch}
\end{figure}

To have the net change in exchange interaction between each ion and its neighbours, we summed all the exchanges and fit the the data with a linear equation. Fig.~\ref{Fig:Tot_exch} shows the change in this net superexchange for the A$_g$(1) and A$_g$(2) modes. As can be seen, the A$_g$(2) mode has a slightly larger slope compared to the A$g$(1) mode and if we include the larger coupling of this mode to the B$_{1u}$(9) mode we could say that this mode has bigger effect in modifying the interaction between \ce{Dy} and \ce{Fe}. However, both interactions result in a change with same sign and the final change in the superexchange will be the sum of the two contributions.

As stated before, the A$_g$(1) mode is dominated by cubic X-point antipolar motions of \ce{Dy} while the A$_g$(2) mode is dominated mainly by the cubic R-point motions of \ce{Dy} which might play a role in making the exchange interactions more sensitive to this mode.
We thus conclude that it is the A$_g$(2) mode excitation which modifies the \ce{Fe}-\ce{Dy} superexchange in the \ce{DyFeO3} dynamics the most and that it can be a driving force of the light-induced renormalization of the magnetic anisotropy potential.

%\textbf{when calculating effect of f electrons, we break the inversion center symmetry and the Dy ions move slightly in C direction and the distance between each of the 8 Fe's and the Dy atom would be different from one another so as can be seen from the exchanges we have 8 different exchanges with different magnitudes and the changes would be with different rates(we have 4 interaction that are stronger compared to other 4, this difference is due to R point antipolar motions of the Dy and then we have X point antipolar motions that splits the 4 interactions into two pairs which results in 4 pairs of interaction and in this case the antipolar motions would result in zero change in superexchange but due to polar motion of the Dy we would have 8 different superexchanges)which results in change in total exchange. \\we also should consider oxygen octahedral motion, because these also change the superexchange path between Fe and Dy and could change the superexchange interactions and in this case I think these motion are important.}

\subsection{Experimental indications of nonlinear lattice dynamics}
\setcounter{subsection}{0}

\begin{figure*}[h]
	\centering
	\includegraphics[width=\linewidth ,keepaspectratio=true]{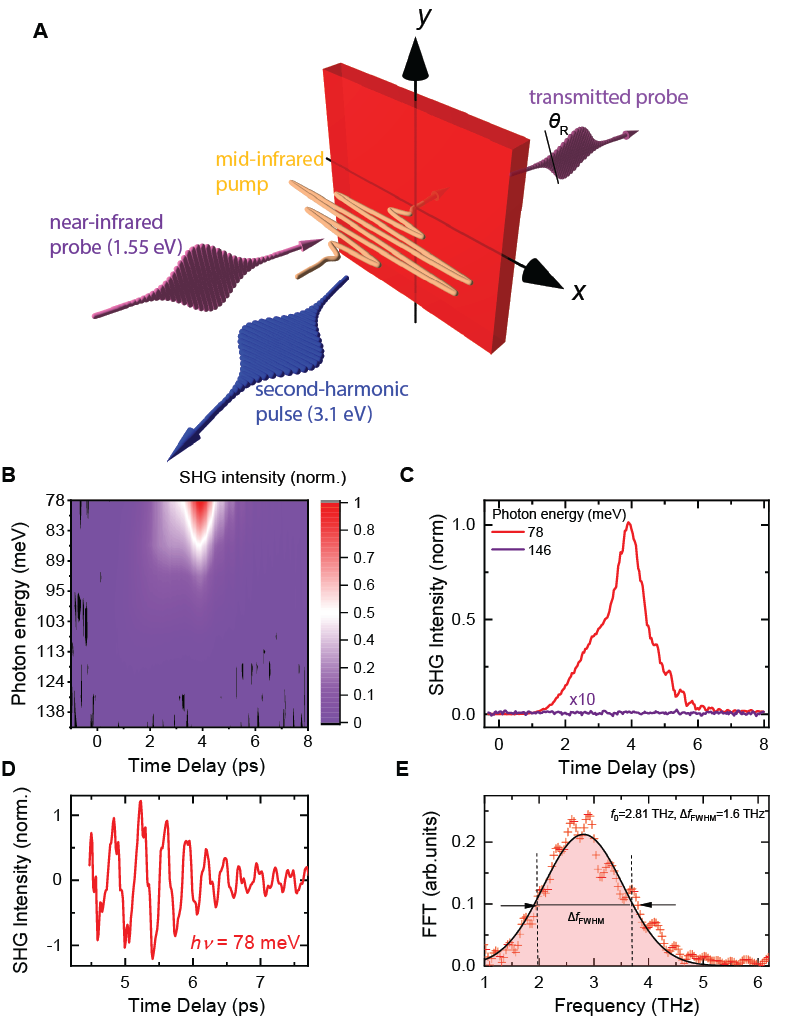} 
	\caption{\textbf{Transient second harmonic generation.} (\textbf{A}) Schematic of the experimental setup. The second harmonic signal generated by the probe pulse after mid-infrared excitation is measured simultaneously with the transmitted probe (Faraday rotaton). (\textbf{B}) Transient second harmonic signal after excitation at different pump energies. (\textbf{C}) Transient second harmonic generation after off-resonant (\SI{146}{\milli\electronvolt}) and resonant (\SI{78}{\milli\electronvolt}) excitation. \textbf{(D)} The oscillations in the second harmonic signal appearing after \SI{4}{\pico\second}, after subtracting the background. \textbf{(E)} Fourier spectrum of the oscillations in the second harmonic signal in (D). The thick solid line is a Gaussian fit. 
	}
	\label{Fig:SHG}
\end{figure*}

In order to gain insight in the structural dynamics induced after phonon excitation, the transient state of the sample was monitored through time-resolved second harmonic generation (SHG) of the probe pulse at the photon energy of \SI{1.55}{\electronvolt} in a reflection geometry. The SHG signal at a photon energy twice the original (\SI{3.1}{\electronvolt}) was measured simultaneously with the rotation of the probe polarization plane in the transmission (Faraday geometry), as is schematically shown in Fig.~\ref{Fig:SHG}A.
In equilibrium, \ce{DyFeO3} has a centrosymmetric structure at all temperatures and therefore no second harmonic signal is observed. However, as the pump pulse is tuned in resonance with the phonon mode, a transient, second harmonic signal appears, as shown in Fig.~\ref{Fig:SHG}B. The strongly wavelength-dependent signal is not observed when the pump pulse energy is off-resonant with the phonon-mode. Examples of these traces are shown in Fig.~\ref{Fig:SHG}C comparing on-resonant  and off-resonant excitation. The wavelength-dependence implies that the second harmonic signal is phonon-driven and indicates the strong connection between phonon-excitation and the transient state of the lattice. An important note must be made that this signal had a strong dependence on the relative orientations of the pump and probe pulse polarization. The largest SHG signal was measured with orthogonal pump and probe polarization (and the measured component of the second harmonic signal parallel to the pump polarization). Compared to the polarization rotation of the transmitted probe (which was used to determine the time zero overlap with the pump), the signal shows a non-trivial delay, followed by a slow rise, peaking after around \SI{4}{\pico\second}. Upon decay, oscillations in this signal appear after \SI{5}{\pico\second}, which are shown in Fig.\ref{Fig:SHG}D. The FFT spectrum  of this oscillatory signal is shown in Fig.~\ref{Fig:SHG}E. The oscillations have a central frequency of \SI{2.8}{\tera\hertz} (\SI{11.6}{\milli\electronvolt}), close to the lowest-energy A$_\textnormal{g}$(1) and B$_\textnormal{3u}$(1) mode (see Table~\ref{Raman active modes} and Table~\ref{IR active modes}). This is a strong indication that, following the concept of nonlinear phononics, excitation of the highest-energy lattice vibrations leads dynamics of lower energy modes.

\FloatBarrier
\newpage

\section*{S5. Ultrafast phonon-driven magnetic phase transition in $\text{DyFeO}_3$}
\setcounter{subsection}{0}

\subsection{Magnetic component of the transient dynamics} 

\begin{figure}[h]
	\centering
	\includegraphics[width=0.9\linewidth]{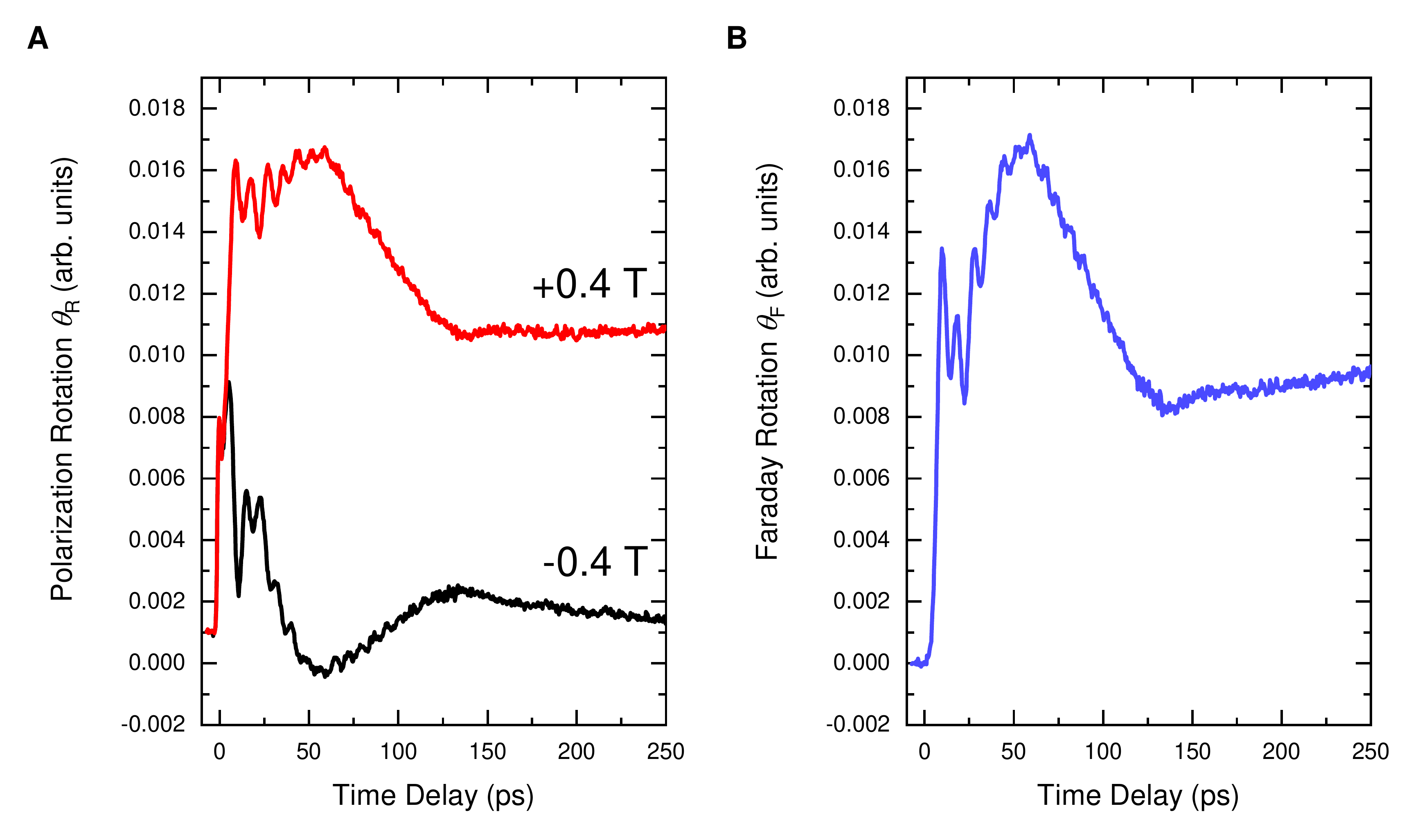}
	\caption{\textbf{Transient magnetic dynamics after phonon excitation.}  (\textbf{A}) Transient polarization rotation as a function of delay time between the pump excitation (\SI{85}{\milli\electronvolt}) and the probe for two applied fields with opposity polarity ($\pm$ \SI{0.4}{\tesla}) at the maximum fluence of \SI{15.8}{\milli\joule\per\centi\meter\squared}. (\textbf{B}) Resulting Faraday rotation after subtracting the traces in (A) according to Eq.~\ref{Eq:Faraday}}
	\label{Fig:S12_PhaseTransition_BothFields}
\end{figure}

In the main text in Fig.~4, the Faraday rotation after phonon excitation is presented for a set of progressive excitation fluences. To avoid an influence of possible artefacts of nonmagnetic origin, we performed the measurements of polarization rotation $\theta_\textnormal{R}$ in vicinity of the Morin transition at two polarities of the applied magnetic field $H$. The Faraday rotation $\theta_\textnormal{F}$ was then calculated as the difference

\begin{equation} 
\theta_\textnormal{F} = \left[\theta_\textnormal{R}(H^+) - \theta_\textnormal{R}(H^-)\right].
\label{Eq:Faraday}
\end{equation} 
This is exemplified in Fig.~\ref{Fig:S12_PhaseTransition_BothFields}. It is this resulting Faraday rotation which was presented in Fig.~4 in the main text.

\subsection{Temperature and field dependence}	
	\begin{figure}[h]
		\centering
		\includegraphics[width =\linewidth]{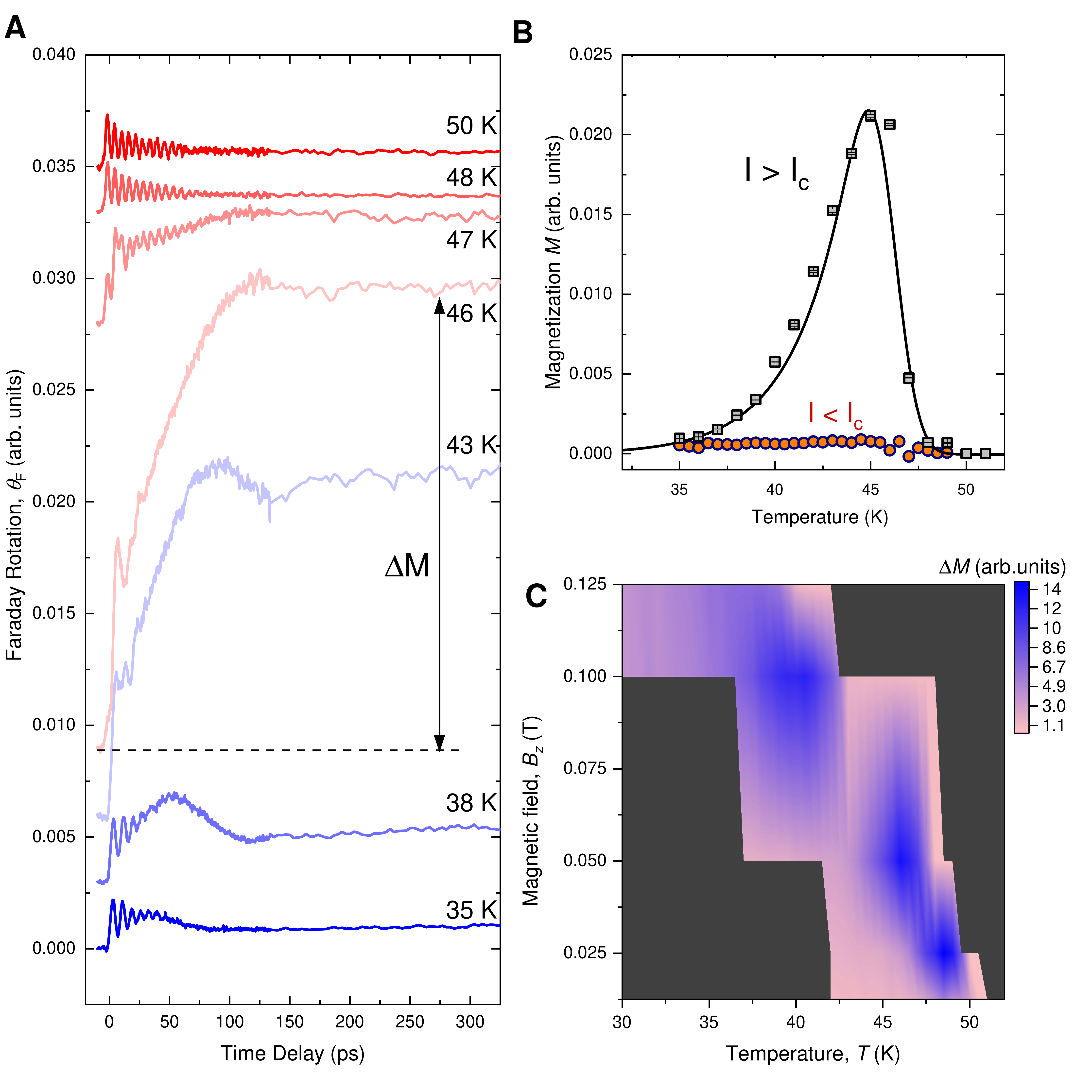}
		\caption{\textbf{Ultrafast phonon-driven magnetic phase transition in \ce{DyFeO3}.} (\textbf{A}) Time-resolved dynamics of the Faraday rotation after resonant excitation (\SI{85}{\milli\electronvolt}) as a function of the sample temperature across the Morin transition for the pump fluence above the threshold $I_\textnormal{c}$. (\textbf{B}) Induced magnetization after resonant excitation as extracted from the data in (A) at $t$=\SI{275}{\pico\second} ($I>I_\textnormal{c}$) and from a similar dataset with lower fluence ($I<I_\textnormal{c}$). (\textbf{C}) The maximal value of the photo-induced magnetization as a function of the temperature and magnetic field.}
		\label{FigureS8}
	\end{figure}
	
Figure \ref{FigureS8}a shows the magnetization dynamics after resonant phonon excitation for different temperatures for a pump fluence $I$ which exceeds the threshold value $I_\textnormal{c}$, $I>I_\textnormal{c}$. A substantial increase in the magnetic signal close to the transition is observed. It is this offset at long time delay vs. temperature $T$ which is shown in Fig.~\ref{FigureS8}B. Similar measurements at a pump fluence below the threshold value led to the other data in this figure. 
The pump-probe measurements in which the phase transition from the AFM to the WFM state was studied, were all performed using a magnetic field. The field is necessary to lift the degeneracy of the WFM states having magnetization directed either parallel or antiparallel with the $z$-axis. The applied field has an influence on the HT-region in which the transient phase transition is observed. This is shown in Fig.~\ref{FigureS8}C, where the magnitude of the photo-induced magnetization is summarized for different applied magnetic fields and temperatures. The transition region shifts to lower temperatures with larger applied field, resembling the behaviour of the Morin temperature $T_\textnormal{M}$ in equilibrium (see Fig.~\ref{Fig:EquiPotential}A). %Moreover, the transition region seems to be broaden for larger applied fields.  

\subsection{Non-thermal kinetics of the phonon-induced magneticc phase changes.}

\begin{figure}[h]
	\centering
	\includegraphics[width=\linewidth]{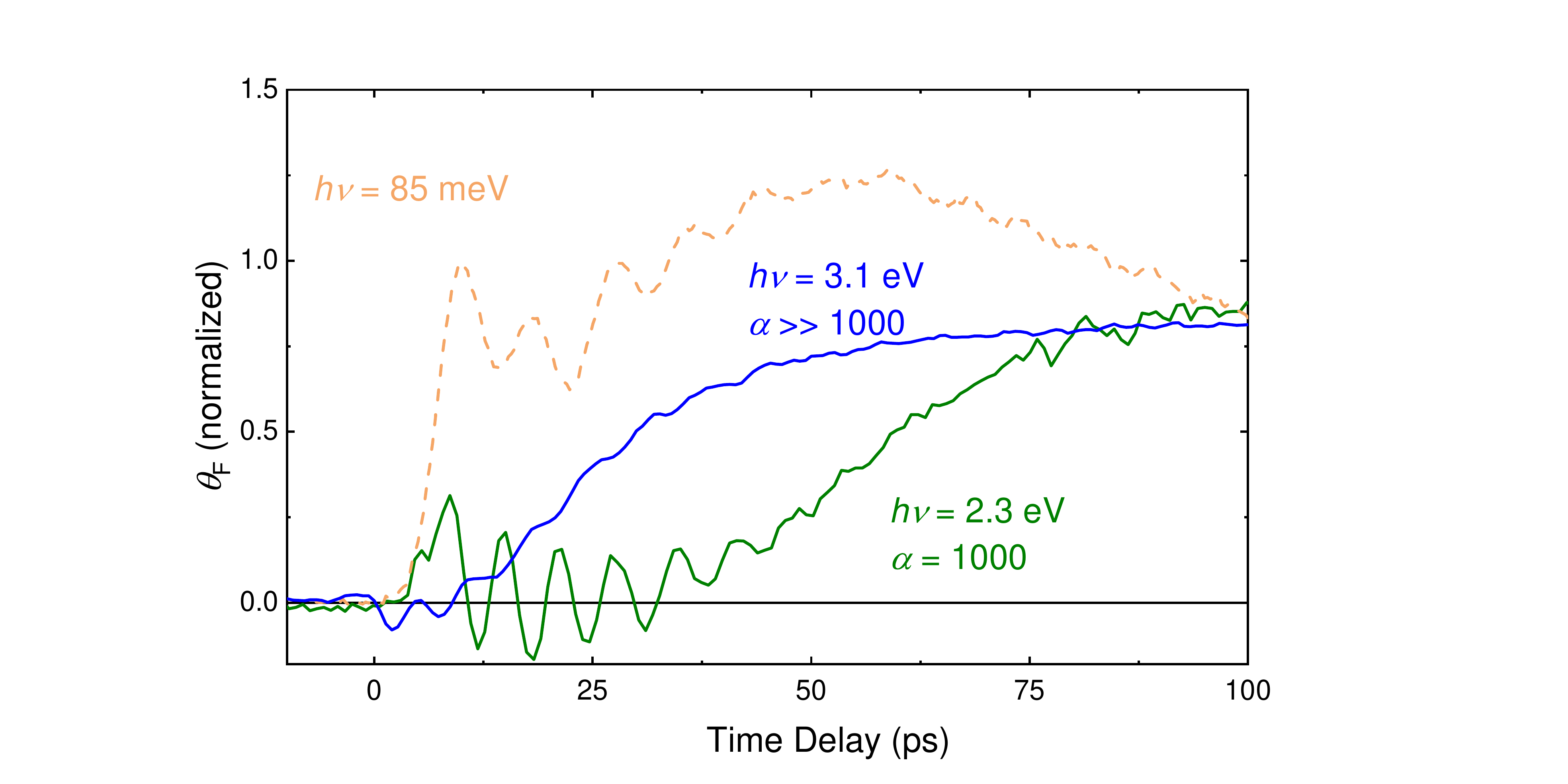}
	\caption{\textbf{Heat-induced vs. phonon-driven magnetic phase transition in \ce{DyFeO3}.} Faraday rotation of the probe pulse ($H = \pm$ \SI{0.3}{\tesla}) after above-bandgap excitation with a pump pulse with photon energy of \SI{2.3}{\electronvolt} (green trace) and \SI{3.1}{\electronvolt} (blue trace). To compare, the kinetics of phonon-induced phase transition after excitation at \SI{85}{\milli\electronvolt} are shown (orange dashed trace)}.
	\label{Fig:S11_500nm_PhaseTransition}
\end{figure}

It is important to compare the observed kinetics of the ultrafast Morin phase transition launched via the phonon-pumping with those driven via a dissipative mechanism. Therefore we performed an auxiliary experiment in which the sample was excited with visible light having a photon energy of \SI{2.3}{\electronvolt} and \SI{3.1}{\electronvolt}, above the change-transfer electronic gap~\cite{Wood1970}. Here the optical absorption is large~($\geq$\SI[scientific-notation = true, per-mode = reciprocal ]{E3}{\per\centi\meter}), see Fig.~\ref{Fig:DFO_absorption} and the transition is expected to be driven by the heat deposited in the medium. As was shown in the main text, Fig.~4D, the experiment shows that although the spin oscillations are launched nearly instantaneously for high photon energy excitation with pump fluences similar to those for the phonon-pumping, it takes more than \SI{30}{\pico\second} before a photo-induced magnetization $\Delta M$ starts to appear and about \SI{100}{\pico\second} to complete the growth. This is shown again in Fig.~\ref{Fig:S11_500nm_PhaseTransition} and compared to the phase transition  dynamics after excitation of the phonon mode. This indicates that the heat-driven evolution of the potential is significantly slower than the one guided by coherent pumping of the lattice. Note that the optical absorption at the above-bandgap energy of \SI{2.3}{\electronvolt} and especially \SI{3.1}{\electronvolt} is significantly larger than the optical absorption due to phonons at \SI{85}{\milli\electronvolt}.
The relatively slow dynamics can be explained by the slow heat transfer from the excited electrons to the $4f$ electrons of \ce{Dy^{3+}}, which determine the dynamics of the magnetic potential~\cite{DeJong2011}. This kinetics stands in a sharp contrast with the one studied here by phonon pumping where the potential is altered on a timescale shorter than a single-cycle of the coherent spin precession. All this leads us to conclude that a very different mechanism from regular heating is at play in the case of phonon pumping.

\FloatBarrier

\bibliographystyle{unsrt_custom}
\bibliography{DyFeO3_v2}
	
\end{document}